\documentclass[aps,prx,twocolumn,superscriptaddress,longbibliography]{revtex4-1}

\usepackage{amsmath}
\usepackage{graphicx}
\usepackage{amsfonts}
\usepackage{verbatim}
\usepackage{amssymb}
\usepackage{color}
\usepackage{dsfont}
\usepackage{amsmath} 
\usepackage{hyperref}

\hypersetup{colorlinks = true, urlcolor = blue, linkcolor = blue, citecolor = blue}

\newcommand{\bmat}{\left(\begin{matrix}}
\newcommand{\emat}{\end{matrix}\right)}
\newcommand{\eabs}{\varepsilon_{\mathrm{A}}}
\newcommand{\eabsm}{\varepsilon_{\mathrm{A},-}}
\newcommand{\eabsp}{\varepsilon_{\mathrm{A},+}}
\newcommand{\eabstilde}{\tilde{\varepsilon}_{\mathrm{A}}}
\newcommand{\eabsmtilde}{\tilde{\varepsilon}_{\mathrm{A},-}}
\newcommand{\eabspmtilde}{\tilde{\varepsilon}_{\mathrm{A},\pm}}
\newcommand{\eabsptilde}{\tilde{\varepsilon}_{\mathrm{A},+}}

\newcommand{\Rtonee}{R^{1\rightarrow 0;e}}

\newcommand{\Rtoneh}{R^{1\rightarrow 0;h}}

\newcommand{\Rtzeroe}{R^{0\rightarrow 1;e}}

\newcommand{\Rtzeroh}{R^{0\rightarrow 1;h}}
\newcommand{\Rtne}{R^{n\rightarrow \bar{n};e}}
\newcommand{\Rtnp}{R^{n\rightarrow \bar{n};p}}
\newcommand{\Rtnh}{R^{n\rightarrow \bar{n};h}}
\newcommand{\Rtnbe}{R^{\bar{n}\rightarrow n;e}}
\newcommand{\Rtnbp}{R^{\bar{n}\rightarrow n;p}}
\newcommand{\Rtnbh}{R^{\bar{n}\rightarrow n;h}}

\newcommand{\Pnq}{P_{q}^{n}}
\newcommand{\Pn}{P^{n}}
\newcommand{\Pnb}{P^{\bar{n}}}
\newcommand{\Pnpqp}{P_{q'}^{\bar{n}}}

\newcommand{\Rnpe}{R^{\bar{n}\rightarrow n;e}_{q'\rightarrow q}}
\newcommand{\Rnph}{R^{\bar{n}\rightarrow n;h}_{q'\rightarrow q}}

\newcommand{\Rne}{R^{n\rightarrow \bar{n};e}_{q\rightarrow q'}}

\newcommand{\Rnh}{R^{n\rightarrow \bar{n};h}_{q\rightarrow q'}}

\newcommand{\Pphq}{P^{\mathrm{b}}_{q}}
\newcommand{\Pphqp}{P^{\mathrm{b}}_{q'}}
\newcommand{\kb}{k_{\mathrm{B}}}

\newcommand{\avex}{\langle \hat{x} \rangle}
\newcommand{\avep}{\langle \hat{p} \rangle}
\newcommand{\Graabs}{G^{r,a}_{\mathrm{A}}}
\newcommand{\Grabs}{G^{r}_{\mathrm{A}}}
\newcommand{\Gaabs}{G^{a}_{\mathrm{A}}}

\newcommand{\Glessabs}{G^{<}_{\mathrm{A}}}
\newcommand{\graabs}{g^{r,a}_{\mathrm{A}}}
\newcommand{\grabs}{g^{r}_{\mathrm{A}}}

\newcommand{\glessabs}{g^{<}_{\mathrm{A}}}

\newcommand{\grlead}{g^{r}_{\mathrm{L}}}

\newcommand{\galead}{g^{a}_{\mathrm{L}}}

\newcommand{\ggreatlead}{g^{>}_{\mathrm{L}}}
\newcommand{\glesslead}{g^{<}_{\mathrm{L}}}

\newcommand{\Sigmaraabs}{\Sigma^{r,a}_{\mathrm{A}}}
\newcommand{\Sigmarabs}{\Sigma^{r}_{\mathrm{A}}}
\newcommand{\Sigmaaabs}{\Sigma^{a}_{\mathrm{A}}}
\newcommand{\Sigmalessabs}{\Sigma^{<}_{\mathrm{A}}}

\newcommand{\Glessleadabs}{G^<_{\mathrm{L}\mathrm{A}}}
\newcommand{\Glessabslead}{G^<_{\mathrm{A}\mathrm{L}}}

\newcommand{\Ggreatabs}{G^>_{\mathrm{A}}}

\newcommand{\Grabsee}{G^{r,ee}_{\mathrm{A}}}
\newcommand{\Grabshh}{G^{r,hh}_{\mathrm{A}}}
\newcommand{\Grabseh}{G^{r,eh}_{\mathrm{A}}}
\newcommand{\Grabshe}{G^{r,he}_{\mathrm{A}}}
\newcommand{\Glessabsee}{G^{<,ee}_{\mathrm{A}}}
\newcommand{\Glessabseh}{G^{<,eh}_{\mathrm{A}}}
\newcommand{\Glessabshe}{G^{<,he}_{\mathrm{A}}}
\newcommand{\Glessabshh}{G^{<,hh}_{\mathrm{A}}}
\newcommand{\Ggreatabshh}{G^{>,hh}_{\mathrm{A}}}
\newcommand{\Ggreatabsee}{G^{>,ee}_{\mathrm{A}}}
\newcommand{\Ggreatabseh}{G^{>,eh}_{\mathrm{A}}}
\newcommand{\Ggreatabshe}{G^{>,he}_{\mathrm{A}}}
\newcommand{\Gaabsee}{G^{a,ee}_{\mathrm{A}}}
\newcommand{\Gaabshh}{G^{a,hh}_{\mathrm{A}}}
\newcommand{\Gaabseh}{G^{a,eh}_{\mathrm{A}}}
\newcommand{\Gaabshe}{G^{a,he}_{\mathrm{A}}}
\newcommand{\ca}{\hat{d}_{\mathrm{A}}}

\newcommand{\cat}{\hat{\tilde{d}}_{\mathrm{A}}}
\newcommand{\NL}{\hat{N}_{\mathrm{L}}}
\newcommand{\dotNL}{\dot{\hat{N}}_{\mathrm{L}}}
\newcommand{\HL}{\hat{H}_{\mathrm{L}}}
\newcommand{\HA}{\hat{H}_{\mathrm{A}}}
\newcommand{\HAtilde}{\hat{\tilde{H}}_{\mathrm{A}}}
\newcommand{\HT}{\hat{H}_{\mathrm{T}}}
\newcommand{\HTtilde}{\hat{\tilde{H}}_{\mathrm{T}}}

\newcommand{\cl}{\hat{c}_{\mathrm{L}}}
\newcommand{\hatclk}{\hat{c}_{\mathrm{L},k}}
\newcommand{\hatcl}{\hat{c}_{\mathrm{L}}}
\newcommand{\hatb}{\hat{b}}
\newcommand{\hattildeb}{\hat{\tilde{b}}}
\newcommand{\hattildegamma}{\hat{\tilde{\gamma}}}
\newcommand{\hatgamma}{\hat{\gamma}}
\newcommand{\hatS}{\hat{S}}
\newcommand{\hatH}{\hat{H}}
\newcommand{\hatA}{\hat{A}}
\newcommand{\hatY}{\hat{Y}}
\newcommand{\hatx}{\hat{x}}
\newcommand{\hatp}{\hat{p}}
\newcommand{\checkt}{\check{t}}
\newcommand{\gravet}{\grave{t}}
\newcommand{\GlessALeepo}{G^{<,ee}_{\mathrm{AL},10}}
\newcommand{\GlessALhepo}{G^{<,he}_{\mathrm{AL},10}}
\newcommand{\GrAeepp}{G^{r,ee}_{\mathrm{A},11}}
\newcommand{\GrAehpp}{G^{r,eh}_{\mathrm{A},11}}
\newcommand{\GrAhepp}{G^{r,he}_{\mathrm{A},11}}
\newcommand{\GrAhhpp}{G^{r,hh}_{\mathrm{A},11}}
\newcommand{\GaAeepp}{G^{a,ee}_{\mathrm{A},11}}
\newcommand{\GaAehpp}{G^{a,eh}_{\mathrm{A},11}}
\newcommand{\GaAhepp}{G^{a,he}_{\mathrm{A},11}}
\newcommand{\GaAhhpp}{G^{a,hh}_{\mathrm{A},11}}
\newcommand{\GlessAeepp}{G^{<,ee}_{\mathrm{A},11}}
\newcommand{\GlessAehpp}{G^{<,eh}_{\mathrm{A},11}}
\newcommand{\GlessAhepp}{G^{<,he}_{\mathrm{A},11}}
\newcommand{\GlessAhhpp}{G^{<,hh}_{\mathrm{A},11}}
\newcommand{\GrAeemm}{G^{r,ee}_{\mathrm{A},-1,-1}}
\newcommand{\GrAehmm}{G^{r,eh}_{\mathrm{A},-1,-1}}
\newcommand{\GrAhemm}{G^{r,he}_{\mathrm{A},-1,-1}}
\newcommand{\GrAhhmm}{G^{r,hh}_{\mathrm{A},-1,-1}}
\newcommand{\GaAeemm}{G^{a,ee}_{\mathrm{A},-1,-1}}
\newcommand{\GaAehmm}{G^{a,eh}_{\mathrm{A},-1,-1}}
\newcommand{\GaAhemm}{G^{a,he}_{\mathrm{A},-1,-1}}
\newcommand{\GaAhhmm}{G^{a,hh}_{\mathrm{A},-1,-1}}
\newcommand{\GlessAeemm}{G^{<,ee}_{\mathrm{A},-1,-1}}
\newcommand{\GlessAehmm}{G^{<,eh}_{\mathrm{A},-1,-1}}
\newcommand{\GlessAhemm}{G^{<,he}_{\mathrm{A},-1,-1}}
\newcommand{\GlessAhhmm}{G^{<,hh}_{\mathrm{A},-1,-1}}
\newcommand{\glessLeeoo}{g^{<,ee}_{\mathrm{L},00}}
\newcommand{\gaLeeoo}{g^{a,ee}_{\mathrm{L},00}}
\newcommand{\grLeeoo}{g^{r,ee}_{\mathrm{L},00}}
\newcommand{\glessLhhoo}{g^{<,hh}_{\mathrm{L},00}}
\newcommand{\gaLhhoo}{g^{a,hh}_{\mathrm{L},00}}
\newcommand{\grLhhoo}{g^{r,hh}_{\mathrm{L},00}}
\newcommand{\GlessALhhmo}{G^{<,hh}_{\mathrm{AL},-1,0}}
\newcommand{\GlessALehmo}{G^{<,eh}_{\mathrm{AL},-1,0}}
\newcommand{\GlessLAeeop}{G^{<,ee}_{\mathrm{LA},01}}
\newcommand{\GlessLAehop}{G^{<,eh}_{\mathrm{LA},01}}
\newcommand{\GlessLAheom}{G^{<,he}_{\mathrm{LA},0,-1}}
\newcommand{\GlessLAhhom}{G^{<,hh}_{\mathrm{LA},0,-1}}
\newcommand{\expomegap}{e^{i\omega_0\tau}}
\newcommand{\expomegam}{e^{-i\omega_0\tau}}
\newcommand{\glessgreatlead}{g_{\mathrm{L}}^{<,>}}
\newcommand{\Glessgreatabs}{G_{\mathrm{A}}^{<,>}}
\newcommand{\lambdatc}{\tilde{\lambda}^{(c)}}
\newcommand{\lambdatd}{\tilde{\lambda}^{(d)}}
\newcommand{\Gttt}{G_{22}^{\mathbb{T}}}

\newcommand{\Gaat}{G_{\alpha\alpha}^{\mathbb{T}}}
\newcommand{\Gaa}{G_{\alpha\alpha}}
\newcommand{\ttilde}{\tilde{t}}
\newcommand{\utilde}{\tilde{u}}
\newcommand{\vtilde}{\tilde{v}}
\newcommand{\elk}{\varepsilon_{\mathrm{L},k}}
\newcommand{\gbar}{\bar{\gamma}}
\newcommand{\bbar}{\bar{b}}
\newcommand{\clkbar}{\bar{c}_{\mathrm{L},k}}
\newcommand{\clbar}{\bar{c}_{\mathrm{L}}}
\newcommand{\clk}{c_{\mathrm{L},k}}

\begin{document}

\title{Electron-boson-interaction induced particle-hole symmetry breaking of conductance into subgap states in superconductors}

\author{F. Setiawan}\email{setiawan@uchicago.edu}
\affiliation{Pritzker School of Molecular Engineering, University of Chicago, 5640 South Ellis Avenue, Chicago, Illinois 60637, USA}
\affiliation{Condensed Matter Theory Center and Joint Quantum Institute, Department of Physics, University of Maryland, College Park, Maryland 20742, USA}
\author{Jay D. Sau}
\affiliation{Condensed Matter Theory Center and Joint Quantum Institute, Department of Physics, University of Maryland, College Park, Maryland 20742, USA}
\date{\today}

\begin{abstract}
Particle-hole symmetry (PHS) of conductance into subgap states in superconductors is a fundamental 
consequence of a noninteracting mean-field theory of superconductivity. The breaking of this PHS
has been attributed to a noninteracting mechanism, i.e., quasiparticle poisoning (QP), a process detrimental to the coherence of superconductor-based qubits.
Here, we show that the ubiquitous electron-boson interactions in superconductors can also break the PHS of subgap conductances. We study the effect of such couplings on the PHS of subgap conductances in superconductors using both the rate equation and Keldysh formalism,
which have different regimes of validity. In both regimes, we found that such couplings give rise to a particle-hole \textit{asymmetry} in subgap conductances which increases with increasing coupling strength, increasing subgap-state particle-hole content imbalance and decreasing temperature. 
Our proposed mechanism is general and applies even for experiments where the subgap-conductance PHS breaking cannot be attributed to QP.

\end{abstract}

\maketitle
\section{Introduction}
Subgap states in superconductors are key features of topological superconducting 
phases~\cite{Mourik2012Signatures,nadj2014observation,Suominen2017Zero,Nichele2017Scaling,menard2017two,choi2017mapping,gul2018ballistic,Deng2018Nonlocality,fornieri2019evidence,ren2019topological,vaitiekenas2020flux,wang2020evidence,zhang2021large} which offer great promise for quantum information processing~\cite{kitaev2003fault,Nayak2008nonabelian}.
Tunneling transport into such Andreev bound states (ABSs) provides the most direct and commonly employed method to detect them~\cite{Law2009Majorana,Flensberg2010Tunneling,wimmer2011quantum,Setiawan2015Conductance} (Hereafter ABS refers to any subgap state in superconductors.) Most of our understanding of tunneling into superconductors is based on the celebrated Blonder-Tinkham-Klapwijk (BTK) formalism~\cite{Blonder1982Transition}. One universal consequence of this theory is a precise particle-hole symmetry (PHS) of the conductance
into any ABS in a superconductor~\cite{Lesovik1997Nonlinearity,wimmer2011quantum,Martin2014Nonequilibrium}. Specifically, this theory predicts that the differential conductance at a positive voltage $V$ inside the superconducting gap precisely match its counterpart value at $-V$. This symmetry has been shown to be a consequence of the PHS of the mean-field Hamiltonian used in the BTK formalism. However, numerous experiments over two decades~\cite{yazdani1997probing,matsuba2003ordered,shan2011observation,Hanaguri2012Scanning,Suominen2017Zero,Nichele2017Scaling,menard2017two,choi2017mapping,gul2018ballistic,Deng2018Nonlocality,chen2018discrete,Bommer2019Spin,Chen2019Ubiquitous,vaitiekenas2020flux,yu2020non,carlos2020temperature,farinacci2020interfering,Spin2021Wang,Ding2021Tuning} have often observed particle-hole (PH) \textit{asymmetric} subgap conductances.  
One way to reconcile this PH asymmetry with the BTK theory is to introduce quasiparticle poisoning induced either by coupling the ABS to a fermionic bath~\cite{Martin2014Nonequilibrium,Nag2015How,Liu2017Role} or through a relaxation process from
the ABS to the superconductor's quasiparticle continuum~\cite{Ruby2015Tunneling}. 

Quasiparticle poisoning (QP)~\cite{Aumentado2004Nonequilibrium,Higginbotham2015Parity,Albrecht2017Transport} refers to a process where an electron tunnels from the bulk of the superconductor to an ABS which changes the occupation (parity) of the ABS. Since the parity is used as the qubit state, 
QP then introduces bit-flip errors~\cite{Goldstein2011Decay,Rainis2012Majorana,Budich2012Failure}. Moreover, as QP breaks the PHS of subgap conductances~\cite{Martin2014Nonequilibrium,Nag2015How,Liu2017Role}, one may be tempted to associate the PH asymmetry to short qubit lifetime. We will show that this correlation is not true in general as contrary to commonly held beliefs, the PH asymmetry can also arise without QP. 


In this paper, we propose a generic mechanism for PHS breaking of subgap conductances without changing the superconductor's parity state, namely, the coupling between ABSs and bosonic modes. While quantum tunneling in dissipative systems has been widely studied~\cite{caldeira1983quantum,ingold1992charge}, previous works consider coupling between bosonic baths and superconductors \textit{without} ABSs. Motivated by tunneling experiments into ABSs~\cite{yazdani1997probing,matsuba2003ordered,shan2011observation,Hanaguri2012Scanning,Suominen2017Zero,Nichele2017Scaling,menard2017two,choi2017mapping,gul2018ballistic,Deng2018Nonlocality,chen2018discrete,Bommer2019Spin,Chen2019Ubiquitous,vaitiekenas2020flux,yu2020non,carlos2020temperature,farinacci2020interfering,Spin2021Wang,Ding2021Tuning}, here we study tunneling transport from a normal lead into an ABS coupled to bosonic modes, e.g., phonons~\cite{Shapiro1975Measurements,Friedl1990Determination}, plasmons~\cite{hepting2018three}, or electromagnetic fields~\cite{majer2007coupling}, in the superconductor. Our system has a local fermion parity analogous to the spin-boson model~\cite{Leggett1987Dynamics} with a caveat that our ABSs can participate in transport. Crucially, our study of transport into an ABS coupled to bosonic modes and its relation to PHS breaking of subgap conductances has not been undertaken before.  To this end, we present ways to enforce fermion-parity conservation in treating interaction effects on transport into  ABSs. We consider two different limits: weak and strong tunneling regimes where the ABS-lead tunnel strength is 
smaller and larger than the thermal broadening $\sim k_B T$, respectively. The weak tunneling limit is studied using the rate equation~\cite{Mitra2004Phonon,Koch2004Thermo}, which is valid for all values of ABS-boson coupling strength where the tunneling rates are calculated using Fermi's Golden Rule (FGR).  In the strong tunneling limit, we study the transport using the Keldysh formalism and treat the ABS-boson coupling within the mean-field approximation.

\section{Particle-Hole symmetry/asymmetry from Fermi's Golden Rule}

\begin{figure}[h!]
\centering
\includegraphics[width=\linewidth]{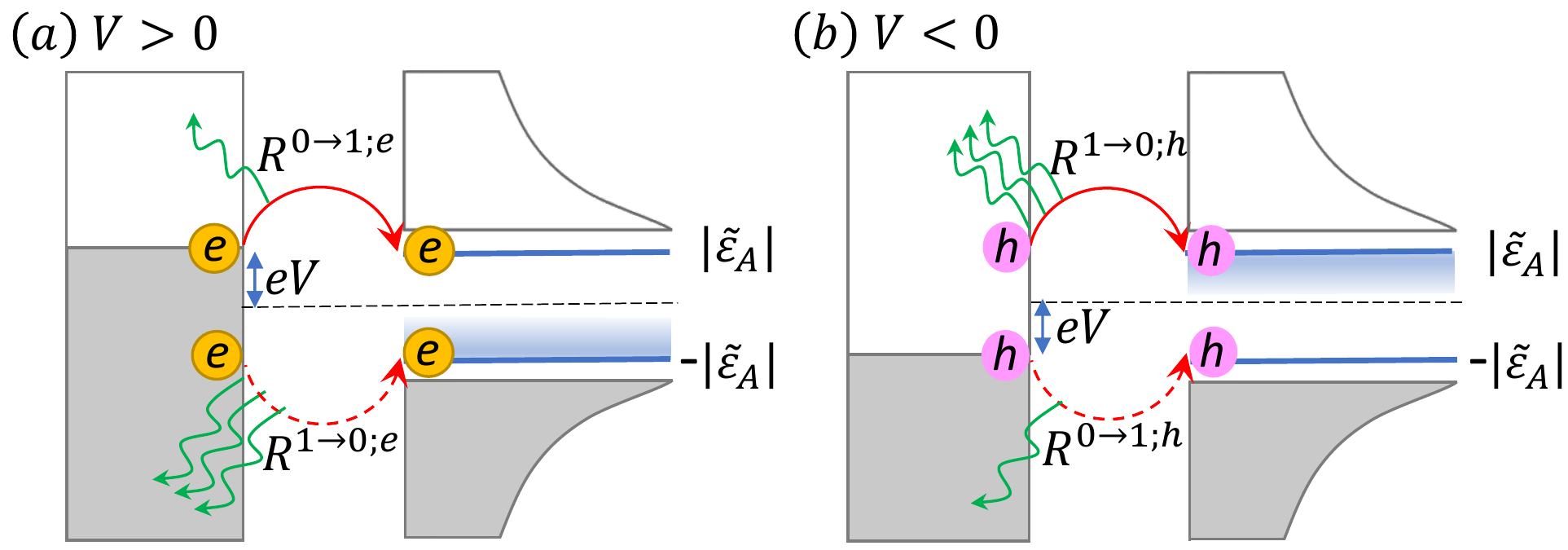}
\caption{Sequential tunneling from the lead (grey rectangles) to the ABS (blue lines). Tunneling of (a) electrons and (b) holes give rise to current at (a) positive and (b) negative voltages, respectively [Eq.~\eqref{eq:currentrate2}]. The first tunneling flips the ABS occupancy $n$ from $0 \rightarrow 1$ and occurs with rates (a) $\Rtzeroe$ or (b) $\Rtzeroh$.  The second tunneling, which flips $n$ from $1 \rightarrow 0$, occurs with rates (a) $\Rtonee$ or (b) $\Rtoneh$. Without bosonic baths, $\Rtzeroe = \Rtoneh$ and $\Rtonee = \Rtzeroh$ giving a PH symmetric conductance. However, in the presence of bosonic baths, 
 the second tunneling occurs with a higher rate since it can transfer lead electrons within a larger energy range near the ABS energy (shaded blue region) where the energy difference can be dumped by emitting bosons (green squiggly lines). Therefore, $\Rtzeroe \neq \Rtoneh$ and $\Rtonee \neq \Rtzeroh$ resulting in a PH asymmetric conductance. 
}\label{fig:schematic}
\end{figure}

We begin by using FGR to show that while subgap conductances in gapped superconductors (superconductors without baths) preserve PHS even with interactions (including strongly correlated superconductors), the PHS is broken for superconductors with gapless excitations (e.g.,  phonons, quasiparticles, etc.). The simplest application of FGR~\cite{yazdani1997probing,mahan2000many,Balatsky2006Impurity} considers the conductance into an ABS at positive [Fig. \ref{fig:schematic}(a)] and negative subgap energies [Fig. \ref{fig:schematic}(b)] to arise from the tunneling of electrons and holes, respectively, into the ABS (changing the ABS occupancy $n$ from $0\rightarrow 1$). The tunneling rates of electrons [$\Rtzeroe$ in Fig. \ref{fig:schematic}(a)] and holes [$\Rtzeroh$ in Fig. \ref{fig:schematic}(b)] can be calculated from FGR to be proportional to the particle and hole component of the ABS wavefunction, respectively. This suggests that the tunneling conductance into an ABS with different weights of particle and hole component is PH asymmetric. However, this simple argument implicitly assumes the presence of QP~\cite{Martin2014Nonequilibrium}, which empties out the electron from the ABS after each tunneling event such that its occupancy $n$ returns to $n=0$. This implicit assumption can be avoided by taking into account
the change in the ABS occupancy $n= 0,1$ after each tunneling. 

As seen in Fig.~\ref{fig:schematic}(a), the electron tunneling flips $n$ either from $0\rightarrow 1$ (with a rate $\Rtzeroe$) or vice-versa (with 
a rate $\Rtonee$). Since each tunneling event flips $n\rightarrow \bar{n}\equiv 1-n$, a full cycle of transferring a pair of 
electrons returns the occupancy to the initial $|n=0\rangle$ occupancy state. The total time for this process that transfers a
 charge of $2e$ is $\tau=(\Rtzeroe)^{-1}+(\Rtonee)^{-1}$ leading to a current $I=2e/\tau$. Combining this result with the analogous argument for negative voltages [Fig.~\ref{fig:schematic}(b)]
 leads to the expression for the tunneling current (we give a more detailed derivation later):
\begin{align}\label{eq:currentrate2}
I = \begin{cases}
2e \dfrac{\Rtzeroe \Rtonee}{\Rtzeroe + \Rtonee} & \text{ for $eV \gtrsim |\eabstilde| + \kb T$,}\\[1em]
 -2e \dfrac{\Rtzeroh \Rtoneh}{\Rtzeroh + \Rtoneh} & \text{ for $eV \lesssim -(|\eabstilde| +\kb T)$,}
\end{cases}
\end{align}
where $\eabstilde$ is the interaction-renormalized ABS energy, $\kb$ is the Boltzmann constant, and $T$ is the temperature. The constraints on the voltage $V$ in Eq.~\eqref{eq:currentrate2} are needed to separate the electron and hole tunneling shown in Fig.~\ref{fig:schematic}. Using FGR, we calculate the electron and hole tunneling rates as $\Rtne \propto |\langle \bar{n}| \ca^\dagger|n\rangle|^2$ and $\Rtnh \propto |\langle \bar{n}| \ca|n\rangle|^2$, where $\ca^\dagger$ and $\ca$ are the electron and hole creation operators in the ABS, respectively. Since $\Rtzeroe= \Rtoneh$ and $\Rtonee = \Rtzeroh$, the current [Eq.~\eqref{eq:currentrate2}] is antisymmetric $I(V) = -I(-V)$ 
and the corresponding subgap conductance shows PHS for a gapped superconductor even with interactions (including strongly correlated superconductors). However, as shown below, this PHS is broken in the presence of bosonic baths.

\section{Model I. Tunneling into boson-coupled-ABS}\label{sec:direct}
We consider tunneling of electrons or holes from a one-dimensional normal lead into an ABS coupled to bosonic modes (e.g., phonons); see Fig.~\ref{fig:schematic}. The total Hamiltonian comprises the Hamiltonian of a boson-coupled ABS, lead, and tunnel coupling, i.e, $\hat{H} = \HA + \HL + \HT$, where
\begin{subequations}\label{eq:Hamiltonian}
\begin{align} 
\HA &= \eabs \hatgamma^\dagger \hatgamma+ \lambda \hatgamma^\dagger \hatgamma(\hatb^\dagger + \hatb)+\Omega \hatb^\dagger \hatb,\label{eq:HABS}\\
\HL &= \sum_{k}\varepsilon_{\mathrm{L},k} \hatclk^\dagger \hatclk, \label{eq:Hleads}\\
\HT &=  t  \hatcl^\dagger\ca + \mathrm{H.c.}\label{eq:HT}
\end{align}
\end{subequations}
Here, $\eabs $ is the ABS energy, $\hatgamma$ ($\hatgamma^\dagger$) is the Bogoliubov annihilation (creation) operator of the ABS, $\lambda$ is the ABS-boson coupling strength, $\hat{b}$ ($\hat{b}^\dagger$) is the boson annihilation (creation) operator, and $\Omega$ is the boson frequency. The operator $\hatclk$ ($\hatclk^\dagger$) annihilates (creates) the lead electron with momentum $k$ and energy $\varepsilon_{\mathrm{L},k}$. The electron tunneling, represented by the Hamiltonian $\hat{H}_T$~\cite{Balatsky2006Impurity,Ruby2015Tunneling}, occurs with a strength $t$ and involves the electron operator of the lead [$\hatcl^\dagger = \int dk \hatclk^\dagger/(2\pi)$] and ABS ($\ca = u \hatgamma + v\hatgamma^\dagger$~\cite{xsuppl}) where $u \equiv u(x=0)$ and $v\equiv v(x=0)$ are the particle and hole component of the ABS wavefunction at the junction $(x = 0)$. We renormalize the ABS wavefunction such that $|u|^2+ |v|^2 = 1$. The ABS-boson coupling term can be derived from the microscopic electron-boson interaction by projecting it onto the lowest-energy (ABS) sector~(see Sec.~I of Ref.~\cite{suppl}). This term can be eliminated using the Lang-Firsov canonical transformation $\hat{\tilde{H}} = e^{\hatS} \hatH e^{-\hatS}$, where $\hatS= \frac{\lambda}{\Omega} \hatgamma^\dagger \hatgamma (\hatb^\dagger - \hatb)$~\cite{lang1963zh,mahan2000many}, which introduces the renormalization $\eabs \rightarrow \eabstilde = \eabs - \lambda^2/\Omega$, $\hatb \rightarrow \hattildeb= \hatb - \lambda \hatgamma^\dagger \hatgamma/\Omega$, $\hatgamma \rightarrow \hattildegamma = \hatgamma \hatY$ and $\ca \rightarrow\cat = u\hatgamma\hatY + v\hatgamma^\dagger\hatY^\dagger$  with $\hatY = e^{-\lambda(\hatb^\dagger - \hatb)/\Omega}$~(see Sec.~II of Ref.~\cite{suppl}). The operator $\hat{Y}$ is analogous to the operator $e^{-i\hat{\varphi}}$ in Ref.~\cite{ingold1992charge}, through the identification  $\lambda (\hat{b}^\dagger -\hat{b})/\Omega = i\hat{\varphi}$ where $\hat{\varphi}$ is the phase operator of the electromagnetic field used in Ref.~\cite{ingold1992charge}. Therefore, our results apply generally to all bosonic modes including electromagnetic fields and plasmons.  


The current operator is  $\hat{I} =-e\dotNL = ie[\NL,\HTtilde]  = i\frac{e}{\hbar} (t\hatcl^\dagger \cat - \mathrm{H.c.})$ where $\dotNL$ is the time derivative of the lead electron number. The current is proportional to the tunnel coupling strength $\Gamma \equiv 2\pi t^2 \nu_0$ where $\nu_0$ is the density of states at the lead Fermi energy. The ratio $\Gamma/\kb T$ determines two different transport regimes: weak ($\Gamma/\kb T < 1$) and strong ($\Gamma/\kb T > 1$) tunneling regimes.

\subsection{Rate equation}
We first study the weak tunneling limit using the rate equation~\cite{Mitra2004Phonon,Koch2004Thermo}, which applies for all values of $\lambda$. Without the lead coupling, the eigenstates of the ABS-boson system are $|n,q\rangle$ with eigenenergies $E_{n,q} = n \eabstilde+ q\Omega$, where the indices $n = 0,1$ and $q \in \mathbb{Z}_{\geq 0}$ denote the ABS and boson occupation numbers, respectively. The tunneling of electrons and holes from the lead to the ABS introduces transitions between the eigenstates $|n,q\rangle$. If the boson relaxation rate is faster than the tunneling rate $\Gamma/\hbar$ (typically true in experiments~\cite{maisi2014andreev}) such that the bosons acquire the equilibrium distribution $\Pphq = e^{-q\Omega/\kb T}(1-e^{-\Omega/\kb T})$, the probability that the system in the state $|n, q\rangle$ can be factorized as $\Pnq=\Pn\Pphq$. In the steady state, $\Pn$ satisfies the rate equation~(see Sec.~III of Ref.~\cite{suppl}):
\begin{align}\label{eq:dP1}
0&= \frac{\partial \Pn}{\partial t} = P^{\bar{n}} \sum_{p=e,h}\Rtnbp  -P^{n}\sum_{p=e,h}\Rtnp,
 \end{align} 
where the probability flux due to the transition from $|\bar{n}\rangle \rightarrow |n\rangle$ and vice versa cancels each other. These transitions rates can be calculated using FGR as~(see Sec.~III of Ref.~\cite{suppl})
\begin{align}\label{eq:R0e}
\Rtne &= \frac{\Gamma}{\hbar} |\langle \bar{n}| \ca^\dagger|n\rangle|^2 \sum_{q,q'}\Pphq  |Y_{qq'}|^2 f(E_{\bar{n},q'} - E_{n,q} - eV),\nonumber\\
\Rtnh &= \frac{\Gamma}{\hbar}|\langle \bar{n}| \ca |n\rangle|^2\sum_{q,q'}\Pphq|Y_{qq'}|^2f(E_{\bar{n},q'} - E_{n,q} + eV),
\end{align}
where $\langle \bar{n}| \ca^\dagger|n\rangle$  and  $\langle \bar{n}| \ca|n\rangle$ are the bare tunneling matrix elements for electrons and holes, respectively, $Y_{qq'} = \langle q'|e^{\lambda(\hatb^\dagger-\hatb)/\Omega}|q\rangle$ is the boson emission or absorption matrix element~\cite{Mitra2004Phonon} (see Sec.~II of Ref.~\cite{suppl}), and $f(E) = [1+\mathrm{exp}({E/k_{\mathrm{B}}T})]^{-1}$ is the lead Fermi function with $T$ being the temperature. Note that  $|\langle 1| \ca^\dagger|0\rangle|^2 = |\langle 0| \ca|1\rangle|^2 = |u|^2$ and $|\langle 0| \ca^\dagger|1\rangle|^2 = |\langle 1| \ca|0\rangle|^2 = |v|^2$~(see Sec.~II of Ref.~\cite{suppl}). Solving Eq.~\eqref{eq:dP1} together with the normalization condition $P^0 + P^1 = 1$, we obtain $P^0$ and $P^1$. Substituting these probabilities into the current $I = e\sum_{n} \Pn\left(\Rtne- \Rtnh \right)$~\cite{Mitra2004Phonon}, we have
\begin{align}\label{eq:currentrate1}
I &= 2e \frac{\Rtzeroe\Rtonee - \Rtzeroh\Rtoneh }{ \Rtzeroe +\Rtoneh +\Rtonee +  \Rtzeroh }.
\end{align}

We can show that Eq.~\eqref{eq:currentrate1} reduces to Eq.~\eqref{eq:currentrate2} by noting that the hole tunneling is energetically forbidden at large positive voltages  ($\Rtzeroh, \Rtoneh \approx 0$ for $eV \gtrsim |\eabstilde| + \kb T$) and so is the electron tunneling at large negative voltages [$\Rtzeroe, \Rtonee \approx 0$ for $eV \lesssim -(|\eabstilde| +\kb T)$]. While Eq.~\eqref{eq:currentrate2} implies PHS for subgap conductances of gapped superconductors, the inclusion of a 
bosonic bath modifies the tunneling rates in Eq.~\eqref{eq:currentrate2} so as to break the conductance PHS. This PHS breaking can be understood more intuitively in the low-temperature limit as follows. The first tunneling, occurring with rates $\Rtzeroe$ [Fig.~\ref{fig:schematic}(a)] or $\Rtzeroh$ [Fig.~\ref{fig:schematic}(b)], transfers only lead electrons or holes near the lead Fermi energy and  is accompanied by emission of small number of bosons since there are only a few occupied electrons (holes) above (below) the Fermi level. In contrast, the second tunneling, whose rates are $\Rtonee$ [Fig.~\ref{fig:schematic}(a)] or $\Rtoneh$~[Fig.~\ref{fig:schematic}(b)], has a higher probability of boson emission since it transfers electrons and holes with energies deep inside the lead Fermi energy. This means that $\Rtzeroe \neq  \Rtoneh$ and $\Rtonee \neq \Rtzeroh$ for tunneling into ABSs in superconductors with gapless excitations (e.g., phonons) unlike the gapped superconductor case. Therefore, $I(V_0) \neq -I(-V_0)$~[Eq.~\eqref{eq:currentrate2}] and the conductance becomes PH asymmetric, i.e., $\left.\frac{dI}{dV}\right|_{V = V_0} \neq \left.\frac{dI}{dV}\right|_{V = -V_0}$~(see Sec.~IV A. of Ref.~\cite{suppl} for a more general proof which holds even for the high-temperature limit).

\begin{figure}[h!]
\centering
\includegraphics[width=\linewidth]{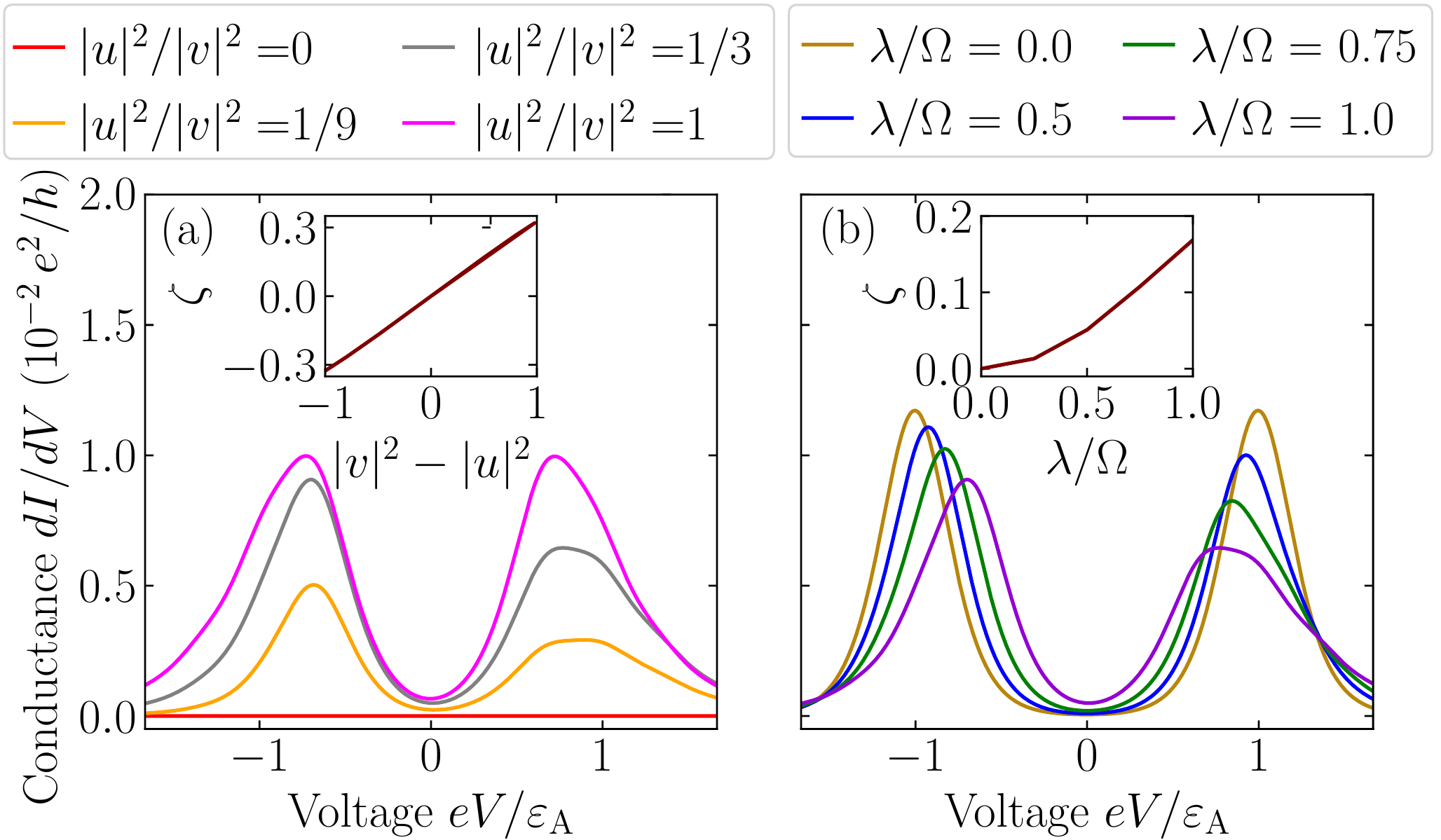}
\caption{Conductance $dI/dV$ of boson-coupled ABSs vs voltage $V$ calculated using the rate equation [Eq.~\eqref{eq:currentrate1}]. Conductances
 for (a) different ratios of PH components $|u|^2/|v|^2$~\cite{zsuppl} with $\lambda/\Omega = 1$ and (b) different ABS-boson coupling strengths $\lambda$ with $|u|^2/|v|^2 = 1/3$.  Inset: (a) Conductance PH asymmetry $\zeta$ vs $|v|^2 - |u|^2$ and (b) $\zeta$ vs $\lambda/\Omega$. Due to the ABS-boson coupling $\lambda$, the ABS energy gets renormalized by $\eabs \rightarrow \eabs -\lambda^2/\Omega$ which shifts the position of the conductance peaks [panel (b)]. The parameters for all panels are: $\eabs/\Omega = 3$, $\Gamma/\Omega = 0.05/(2\pi)$, and $\kb T/\Omega = 0.4$.}\label{fig:currentrate}
\end{figure} 

Figure~\ref{fig:currentrate} shows the conductance $dI/dV$ (see Sec.~V of Ref.~\cite{suppl} for the current) of boson-coupled ABSs calculated from Eq.~\eqref{eq:currentrate1}. As shown in Fig.~\ref{fig:currentrate}(a), the conductance decreases with increasing ABS's PH content imbalance $||u|^2 - |v|^2|$ because the terms $\Rtzeroe \Rtonee$ and $\Rtzeroh \Rtoneh$ in Eq.~\eqref{eq:currentrate1} are $\propto |uv|^2 = [1-(|u|^2 - |v|^2)^2]/4$. In contrast, the conductance PH asymmetry magnitude $|\zeta|$ increases linearly with increasing ABS's PH content imbalance $||u|^2 - |v|^2|$[inset of Fig.~\ref{fig:currentrate}(a)], where 
\begin{equation}
\zeta \equiv \frac{(dI/dV)_{\mathrm{max},-} - (dI/dV)_{\mathrm{max},+}}{(dI/dV)_{\mathrm{max},-} + (dI/dV)_{\mathrm{max},+}},
\end{equation}
with $(dI/dV)_{\mathrm{max},-/+} \equiv \max_{V \leq 0/ V\geq 0} dI/dV$ being the peak conductance at negative and positive voltages, respectively. $\zeta = \pm 1$ ($\zeta = 0$) corresponds to perfectly asymmetric (symmetric) conductances. Figure~\ref{fig:currentrate}(b) shows that the peak conductances decrease with increasing ABS-boson coupling strength $\lambda$ since $\lambda$ broadens the quasiparticle weight around the ABS energy, which decreases the effective tunnel coupling strength. The conductance PH asymmetry ($\zeta$) magnitude~\cite{vsuppl}, however, increases with increasing $\lambda$~[inset of Fig.~\ref{fig:currentrate}(b)] for $|\eabstilde| \gg \kb T$ where the two peaks are well separated. For the regime where $|\eabstilde| \approx \kb T$, $\zeta$ has a nonmonotonic behavior with $\lambda$~(see Sec.~VI of Ref.~\cite{suppl}). Section VI of Ref.~\cite{suppl} shows that $\zeta$ decreases with increasing temperature, depends monotonically on the boson frequency $\Omega$, and prevails only for $\Omega \lesssim  2|\eabstilde| + \kb T$.

\subsection{Keldysh}
For strong-tunneling limit ($\Gamma > \kb T$), we compute the current using the mean-field Keldysh formalism. We begin by rewriting Eq.~\eqref{eq:HABS} in terms of the boson displacement [$\hat{x} = (\hatb + \hatb^\dagger)/\sqrt{2}$] and momentum [$\hat{p} = -i(\hatb - \hatb^\dagger)/\sqrt{2} $] operator as
\begin{align}\label{eq:HABSa}
\HA &= \eabs \hatgamma^\dagger \hatgamma + \sqrt{2}\lambda  \hatgamma^\dagger \hatgamma \hatx + \frac{\Omega}{2}(\hatx^2 + \hatp^2) -\frac{\Omega}{2}.
\end{align}
We calculate the mean-field energy $\eabs + \sqrt{2}\lambda \avex$ by self-consistently solving for $\avex$ where $\langle \cdots \rangle$ is the expectation value with respect to the mean-field eigenfunction. To this end, we solve for $\partial \langle \HA\rangle/\partial\avex = 0$ and $\partial \langle \HA\rangle/\partial\avep = 0$, giving $\avex = -\frac{\sqrt{2}\lambda}{\Omega}\langle \hatgamma^\dagger \hatgamma \rangle$ and  $\avep = 0$. 

The ABS Green's function in the Lehmann representation is $g_\mathrm{A}(\omega) = \frac{\Phi_+ \Phi^\dagger_+}{\omega - (\eabs+\sqrt{2}\lambda \avex)} +  \frac{\Phi_- \Phi^\dagger_-}{\omega + (\eabs+\sqrt{2}\lambda\avex)}$ where $\Phi_+ = (1, 0)^T$ and $\Phi_- = (0, 1)^T$ are the Nambu spinors written in the Nambu basis $(\hatgamma,\hatgamma^\dagger)^T$. Following Ref.~\cite{Ruby2015Tunneling}, we use the Green's function to evaluate the current as (see Sec.~VII of Ref.~\cite{suppl})
\begin{equation}\label{eq:IV}
 I(V) = \frac{e}{h}\Gamma^2\int_{-\infty}^{\infty} d\omega \mathcal{A}(\omega)[f(\omega_-)-f(\omega_+)],
\end{equation}
where
\begin{equation}\label{eq:Aomega}
\mathcal{A}(\omega) = \frac{4 |uv|^2}{\left[\omega -\frac{(\eabs+\sqrt{2}\lambda \avex)^2}{\omega}- \frac{(\Gamma_u - \Gamma_v)^2}{4\omega}\right]^2+ (\Gamma_u + \Gamma_v)^2},
\end{equation}
with $\Gamma_u  = \Gamma |u|^2$ and $\Gamma_v  =  \Gamma |v|^2$. The mean-field boson displacement $\avex$ in Eq.~\eqref{eq:Aomega} is evaluated self-consistently as
\begin{align}\label{eq:selfconst}
\avex &= -\frac{\sqrt{2}\lambda}{ \Omega} \langle \hatgamma^\dagger \hatgamma \rangle = -\frac{\lambda}{\sqrt{2}\Omega}\left\{1-i\int \frac{d\omega}{2\pi} \mathrm{Tr}\left[\Glessabs(\omega)\sigma_z\right]\right\},
\end{align}
where $(G^<_{\alpha \beta})_{\mathrm{A}} = i\langle \Psi^\dagger_{\alpha \mathrm{A}}\Psi_{\beta \mathrm{A}}\rangle$ is the ABS lesser Green's function~(see Sec.~VIII of Ref.~\cite{suppl}) with $\Psi_{\mathrm{A}} = (\hatgamma,\hatgamma^\dagger)^T$  and $\sigma_z$ being the $z$-Pauli matrix in the Nambu space.


\begin{figure}[h!]
\centering
\includegraphics[width=\linewidth]{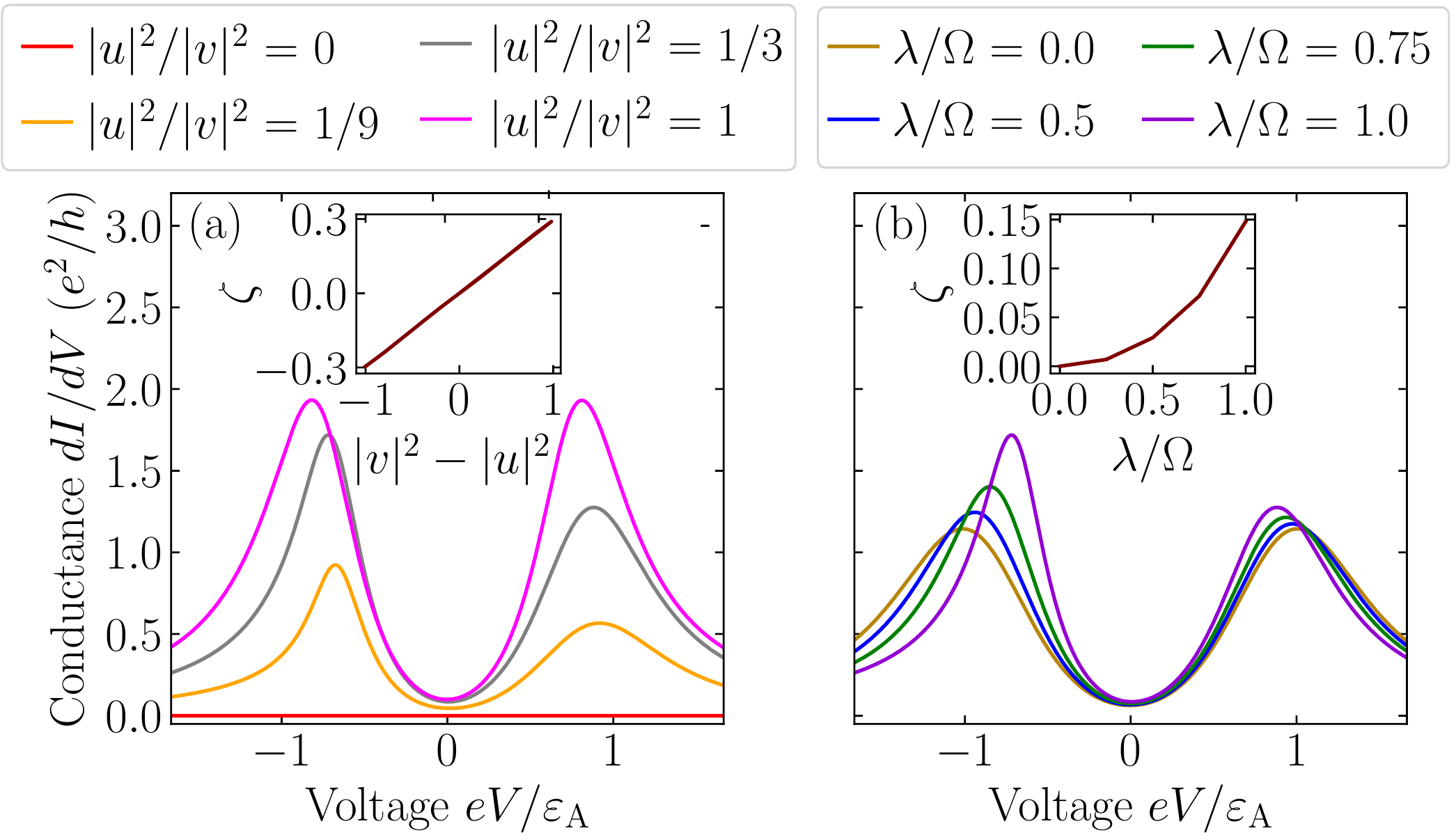}
\caption{Conductance $dI/dV$ of boson-coupled ABSs vs voltage $V$ calculated using the mean-field Keldysh approach [Eq.~\eqref{eq:IV}]. Conductances for (a) different ratios of PH components $|u|^2/|v|^2$~\cite{zsuppl} with $\lambda/\Omega = 1$ and (b) different ABS-boson coupling strengths $\lambda$ with $|u|^2/|v|^2=1/3$. Inset: (a) Conductance PH asymmetry $\zeta$ vs $|v|^2 - |u|^2$ and (b) $\zeta$ vs $\lambda/\Omega$. Due to the ABS-boson coupling $\lambda$, the ABS energy gets renormalized by $\eabs \rightarrow \eabstilde = \eabs +\sqrt{2}\lambda\avex$ (see Sec.~V of Ref.~\cite{suppl} for the plots of $\avex$), which shifts the position of the conductance peak [panel (b)]. The parameters for all panels are: $\eabs/\Omega = 3$, $\Gamma/\Omega = 2$, and $\kb T/\Omega = 0.4$. }\label{fig:Keldyshcurrent}
\end{figure}

Figure~\ref{fig:Keldyshcurrent} shows the conductance (see Sec.~V of Ref.~\cite{suppl} for the current) of boson-coupled ABSs calculated from Eq.~\eqref{eq:IV} subject to the self-consistency condition [Eq.~\eqref{eq:selfconst}]. Similar to the rate equation, the conductance of boson-coupled ABSs  calculated using the Keldysh approach also decreases with increasing ABS's PH content imbalance $||u|^2 - |v|^2|$ [Fig.~\ref{fig:Keldyshcurrent}(a)] with its PH asymmetry ($\zeta$) magnitude increases linearly with increasing $||u|^2 - |v|^2|$ [inset of Fig.~\ref{fig:Keldyshcurrent}(a)].  Figure~\ref{fig:Keldyshcurrent}(b) shows that the peak conductance increases with increasing ABS-boson coupling strength $\lambda$ contrary to the rate-equation results. However, similar to the rate-equation, the conductance PH asymmetry $\zeta$ increases with increasing $\lambda$~\cite{usuppl}. Unlike the rate equation, the Keldysh approach shows that in the strong-tunneling regime the PHS breaking holds also for high-frequency bosons~(see Sec.~IX of Ref.~\cite{suppl}), since it arises from nonperturbative effects of tunneling, i.e., the PH asymmetry of the mean-field boson displacement value $\avex$.

Our model of tunneling into boson-coupled ABS~[Eq.~\eqref{eq:Hamiltonian}] can explain the origin of PH asymmetry for subgap conductance observed in a hard superconducting gap~\cite{menard2017two,choi2017mapping,carlos2020temperature,farinacci2020interfering,Ding2021Tuning} which cannot be accounted for by QP. However, similar to QP this model also results in conductance peak areas which are independent of temperatures~(see Sec.~IV B. of Ref.~\cite{suppl}). In Sec.~\ref{sec:indirect} below, we consider another related model, i.e., a boson-assisted tunneling model. This model can not only give rise to PHS breaking of subgap conductances but also account for experimentally observed conductance features which cannot be attributed to QP, e.g., an increase in the conductance peak area with temperature~\cite{carlos2020temperature}.

\section{Model II. Boson-assisted tunneling into ABS}\label{sec:indirect}
In this section, we consider boson-assisted tunneling into an ABS via virtual hopping of electrons or holes from the lead into higher-lying states in superconductors which are boson-coupled to the ABS. The higher-lying states can be either higher-energy ABSs or states from the continuum above the gap. By integrating out the higher-lying states, we derive the effective low-energy Hamiltonian for the boson-assisted tunneling into the ABS as~(see Sec.~X of Ref.~\cite{suppl})
\begin{align}
\HT&= t(\hatb + \hatb^\dagger)\hatcl^\dagger\ca+\mathrm{H.c.}
\end{align}
Note the extra $(\hatb + \hatb^\dagger)$ term in the above tunneling Hamiltonian as compared to Eq.~\eqref{eq:HT} in Sec.~\ref{sec:direct}. 


\begin{figure}[h!]
\centering
\includegraphics[width=\linewidth]{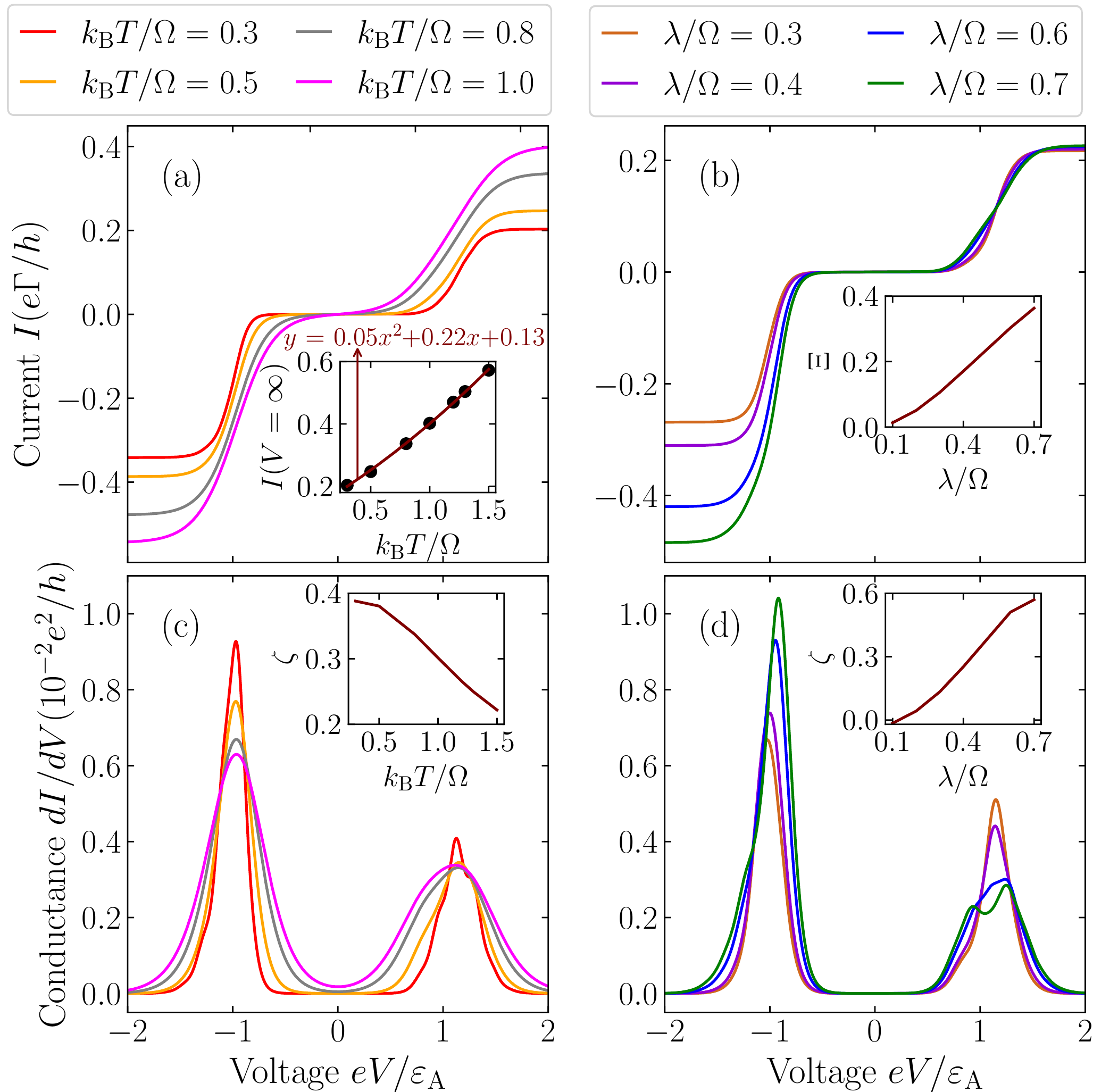}
\caption{Current $I$ and conductance $dI/dV$ into ABS vs voltage $V$ for the boson-assisted tunneling model. (Upper panels) Current and (lower panels) conductance for (left panels) different temperatures $\kb T$ with $\lambda/\Omega = 0.5$ and (right panels) different ABS-boson coupling strengths $\lambda$ with temperature $\kb T = 0.4$.  Inset: (a) $I(V=\infty)$ vs $\kb T/\Omega$, (b) Current PH asymmetry $\Xi$ vs $\lambda/\Omega$, (c) Conductance PH asymmetry $\zeta$ vs $\kb T/\Omega$, and (d) $\zeta$ vs $\lambda/\Omega$. Due to the ABS-boson coupling $\lambda$, the ABS energy gets renormalized by $\eabs \rightarrow \eabs -\lambda^2/\Omega$ which shifts the position of the conductance peaks [panel (d)]. The parameters for all panels are: $|u| = 0.1$, $\eabs/\Omega = 6$, and $\Gamma/\Omega = 0.05/(2\pi)$.}\label{fig:currenttwoABS}
\end{figure}

Figure~\ref{fig:currenttwoABS} shows the current and conductance calculated using the rate equation within the boson-assisted tunneling model for different temperatures $T$ and ABS-boson coupling strengths $\lambda$. Contrary to the tunneling model in Sec.~\ref{sec:direct} where the current at $V = \pm \infty$ is independent of temperature (see Sec.~IV B. of Ref.~\cite{suppl}), for the boson-assisted tunneling model the current magnitude at $V = \pm \infty$ increases with increasing temperature. This is because the boson-assisted tunneling rate~(see Sec.~X of Ref.~\cite{suppl}) is proportional to $\langle q| (\hattildeb + \hattildeb^\dagger)^2 |q\rangle$, which increases with increasing temperature. Crucially, we find that the current  $I(V =\pm \infty)$ or equivalently the peak area of the conductance versus voltage curve has a faster-than-linear increase with temperature [inset of Fig.~\ref{fig:currenttwoABS}(a)], providing excellent agreement with experimental results~\cite{carlos2020temperature}. Since QP preserves the conductance peak area under different temperatures and necessarily induces ``soft-gap" conductance features, our proposed boson-assisted tunneling process is thus more likely to be responsible for the PH-asymmetric subgap conductances inside a hard superconducting gap observed in Ref.~\cite{carlos2020temperature}. 

Contrary to the model in Sec.~\ref{sec:direct}, for the boson-assisted tunneling model, the current calculated at large positive and negative voltages need not be perfectly antisymmetric, i.e., $I(V = \infty) \neq -I(V = -\infty)$. The current PH asymmetry $\Xi$, defined as 
\begin{equation}
\Xi \equiv \frac{|I(V=-\infty)|-|I(V=\infty)|}{|I(V=-\infty)|+|I(V=\infty)|},
\end{equation}
increases with increasing ABS-boson coupling strength $\lambda$ [see inset  of Fig.~\ref{fig:currenttwoABS}(b)]. This current PH asymmetry (or equivalently the asymmetry between the conductance peak area for positive and negative voltages) as well as the dependence of the conductance peak area with temperature can serve as signatures for the boson-assisted tunneling process. Similar to the model in Sec.~\ref{sec:direct}, the conductance PH asymmetry $\zeta$ calculated using the boson-assisted tunneling model also decreases with increasing temperature [inset of Fig.~\ref{fig:currenttwoABS}(c)] and increases with increasing ABS-boson coupling strength $\lambda$ [inset of Fig.~\ref{fig:currenttwoABS}(d)].

\section{Conclusions} 
Contrary to widely held belief, we show that the PHS breaking of subgap conductances in superconductors can arise without QP. Specifically, the coupling of ABSs to a bosonic bath (or multimode bosonic baths~\cite{multimode}) can break the PHS of subgap conductances without changing the superconductor's parity state. Therefore, contrary to QP, our mechanism is not detrimental to the coherence of superconductor-based qubits. (Topological qubits are exponentially protected from the bosonic bath dephasing due to the spatial separation of Majoranas~\cite{Knapp2018Dephasing}.) We find that the conductance PH asymmetry increases with increasing ABS's PH content imbalance, increasing ABS-boson coupling strength and decreasing temperature. Our theory is general as it applies to all ABSs, e.g., quasi-Majorana states~\cite{Kells2012Near,Liu2017Andreev}, Yu-Shiba-Rusinov states~\cite{Yu1965Acta,shiba1968classical,rusinov1969theory}, Caroli-de~Gennes-Matricon states~\cite{caroli1964bound}, etc., which couple to bosonic modes such as phonons, plasmons, electromagnetic fields, etc., in superconductors. Contrary to QP, our mechanism applies even for ABSs observed inside a hard superconducting gap~\cite{menard2017two,choi2017mapping,carlos2020temperature,Ding2021Tuning} and can give rise to an increase in the conductance peak area with temperature as observed in experiments~\cite{carlos2020temperature}. 

Our PHS breaking mechanism results from boson emissions or absorptions accompanying the electron/hole tunneling.  Since these bosons such as phonons are ubiquitous in superconductors, we expect electron-phonon interactions (EPIs) to significantly affect transport in superconductors, particularly the semiconductor-superconductor
heterostructures used to realize topological superconductors~\cite{Suominen2017Zero,Nichele2017Scaling,menard2017two,gul2018ballistic,Deng2018Nonlocality,Bommer2019Spin,Chen2019Ubiquitous,vaitiekenas2020flux,yu2020non}. In fact, measurements  of transport in semiconductors have observed features~\cite{Huntzinger2000Acoustic,Goldman1987Evidence,Hartke2018Microwave,roulleau2011coherent,Weber2010Probing}
 associated with EPI that are theoretically understood~\cite{Dephasing2004Muljarov,Kleinman1965Theory,Wingreen1988Resonant}. 
 We estimate that for a typical topological superconductor which uses either an InAs or InSb semiconductor with a length of $\ell \sim 1$ $\mu$m (having a phonon frequency $\Omega \sim \hbar v_s \pi/\ell = 7.2$ $\mu$eV where $v_s \approx 3.5 \times 10^5$ cm/s~\cite{yano1993raman,wagner1995raman,madelung2012semiconductors} is the sound velocity), EPI can give rise to a conductance PH asymmetry in the tunneling limit for ABSs with energies $\eabs \gtrsim \Omega/2 = 3.6$ $\mu$eV. Therefore, contrary to QP, EPI does not affect the zero-bias Majorana conductance. 

Compared to diagrammatic techniques, FGR is a more controlled approach in treating the effect of interactions 
on transport in superconductors (even for strongly correlated superconductors) for the strict tunneling limit. This is because interaction diagrams 
can generate an \textit{imaginary}  self-energy~\cite{Ruby2015Tunneling},
resulting in a conductance PH asymmetry similar to QP~\cite{Martin2014Nonequilibrium}. Therefore, it is crucial 
to enforce a fermion parity conservation in the diagrammatic treatment of ABS-boson couplings like our mean-field treatment of interactions in the Keldysh formulation. Our work thus motivates the formulation of the nonequilibrium Green's function beyond the mean field approximation that conserves fermion parity. 
We note that our mechanism is quite 
distinct from the subgap-conductance PHS breaking due to the bias-voltage dependence of the tunnel barrier~\cite{WimmerPrivate}. 
 While this mechanism can be treated within the Keldysh approach by moving 
the interaction term from the ABS to the barrier, it vanishes in the tunneling limit where FGR applies.

\acknowledgements
We thank M. Wimmer, J. Saldana and R. Hanai for useful discussions.  This work was supported by Microsoft research, Army Research Office Grant no. W911NF-19-1-0328, NSF DMR-1555135 (CAREER), JQI-NSF-PFC (supported by NSF grant PHY-1607611), and NSF PHY-1748958 (through helpful discussions at KITP).

\bibliography{manuscript}

\begin{thebibliography}{86}
\bibitem [{xsu()}]{xsuppl}%
  \BibitemOpen
  \href@noop {} {}\bibinfo {note} {Since we consider only the subgap state and
  ignore the above-gap states, the relation $\hat{d}_{\mathrm{A}} = u\gamma +
  v\gamma^\dagger$ is only approximate which makes $\hat{d}_{\mathrm{A}}$
  nonfermionic. The operator $\ca$ becomes fermionic if all the states in the
  superconductor including the above-gap states are taken into account [see Eq.
  (S-5) in Sec.~I of Ref.~\cite{suppl}]. Our conclusion on the PHS breaking of
  the subgap conductance due to the ABS-boson coupling does not rely on the
  fermionic properties of $\ca$.}\BibitemShut {Stop}%
\bibitem [{zsu()}]{zsuppl}%
  \BibitemOpen
  \href@noop {} {}\bibinfo {note} {The conductance for $|v|^2 > |u|^2$ can be
  obtained from the conductance for $|v|^2 < |u|^2$ (shown in
  Figs.~\ref{fig:currentrate} and~\ref{fig:Keldyshcurrent}) by interchanging
  both $|u|\leftrightarrow |v|$ and $V \leftrightarrow -V$ simultaneously. As a
  result, the higher and lower peaks switch sides which changes the sign of the
  PH asymmetry $\zeta$ (see Sec.~IV A of Ref.~\cite{suppl} for a
  proof).}\BibitemShut {Stop}%
\bibitem [{usu()}]{usuppl}%
  \BibitemOpen
  \href@noop {} {}\bibinfo {note} {Since we ignore the Fock term in the
  mean-field approximation, the conductance calculated from the Keldysh
  approach has no boson sidebands.}\BibitemShut {Stop}%
\bibitem [{mul()}]{multimode}%
  \BibitemOpen
  \href@noop {} {}\bibinfo {note} {While here we focus exclusively on
  single-mode bosonic baths, our proposed mechanism is expected to hold also
  for multimode bosonic baths since the result for multimode bosonic baths is
  qualitatively similar to averaging multiple results for different
  single-bosonic modes. This averaging is justified as bosonic modes do not
  interact with each other and are therefore independent.}\BibitemShut {Stop}%
\end{thebibliography}

\clearpage

\onecolumngrid
\vspace{1cm}
\begin{center}
{\bf\large Supplemental Material for ``Electron-boson-interaction induced particle-hole symmetry breaking of conductance into subgap states in superconductors"}
\end{center}
\vspace{0.5cm}

\setcounter{secnumdepth}{3}
\setcounter{equation}{0}
\setcounter{figure}{0}
\setcounter{section}{0}
\renewcommand{\theequation}{S-\arabic{equation}}
\renewcommand{\thefigure}{S\arabic{figure}}
\renewcommand\figurename{Supplementary Figure}
\renewcommand\tablename{Supplementary Table}
\newcommand\Scite[1]{[S\citealp{#1}]}
\newcommand\Scit[1]{S\citealp{#1}}

\makeatletter \renewcommand\@biblabel[1]{[S#1]} \makeatother

\section{Derivation of the ABS-boson coupling from the microscopic electron-boson interaction}
In this section, we derive the ABS-boson coupling term in Eq.~\eqref{eq:HABS} of the main text from the microscopic electron-boson interaction. We begin by writing a generic Hamiltonian for a superconductor with an ABS as
\begin{equation}
\hat{H}_{\mathrm{SC}} = \sum_{l,m}\int dx \int dx' h_{lm}(x,x') \hat{d}_{l}^\dagger(x) \hat{d}_{m}(x') + \left[\Delta_{lm}(x,x') \hat{d}_{l}^\dagger(x) \hat{d}^\dagger_{m}(x') + \mathrm{H.c.}\right],
\end{equation}
where $h_{l,m}$ describes the dynamics of the electrons in the superconductor with an ABS, $\Delta_{l,m}$ is the superconducting pairing potential, $d_{l,m}^\dagger$ ($d_{l,m}$) is the electron creation (annihilation) operator of the superconductor and the indices $l,m$ represent both the orbital and spin degrees of freedom. We can diagonalize the above Hamiltonian using the Bogoliubov transformation 
\begin{subequations}
\begin{align}
\hat{\gamma}_\alpha &= \sum_m\int dx \left[u_{\alpha m}^*(x) d_m (x) + v_{\alpha m}(x) d_m^\dagger (x) \right], \\
\hat{\gamma}_\alpha^\dagger &= \sum_m\int dx \left[ v_{\alpha m}^*(x) d_m (x) +u_{\alpha m}(x) d_m^\dagger (x) \right],
\end{align}
\end{subequations}
which gives
\begin{align}
\hat{H}_{\mathrm{SC}} = \sum_{\alpha} \varepsilon_\alpha \hat{\gamma}_\alpha^\dagger \hat{\gamma}_\alpha+\mathrm{const},
\end{align}
where the lowest energy level corresponds to the ABS energy, i.e., $\varepsilon_1 = \eabs$.

The Hamiltonian of the electron-boson coupling is given by
\begin{equation}\label{eq:Heph}
\hat{H}_{\text{e-b}} = \sum_{lm}\int dx \int dx'g_{lm}(x,x') \hat{d}^\dagger_l(x) \hat{d}_m(x') (\hat{b}^\dagger+\hat{b}),
\end{equation}
where $g_{lm}$ is the electron-boson coupling strength, and $\hatb$ ($\hatb^\dagger$) is the boson annihilation (creation) operator.
Substituting 
\begin{subequations}\label{eq:dm}
\begin{align}
d_m(x) &= \sum_{\alpha>0}\left[ u_{m\alpha}(x) \hatgamma_\alpha +v_{m\alpha}(x) \hatgamma_\alpha^\dagger   \right],\label{eq:dm}\\
d_m^\dagger(x) &= \sum_{\alpha>0} \left[  v_{m\alpha}^*(x) \hatgamma_\alpha + u_{m\alpha}^*(x) \hatgamma_\alpha^\dagger \right],
\end{align}
\end{subequations}
into Eq.~\eqref{eq:Heph},
we have 
\begin{align}\label{eq:Hephmicro}
\hat{H}_{\text{e-b}} &= \sum_{lm\alpha\beta}\int dx \int dx'g_{lm}(x,x') \left[v_{l\alpha}^*(x) u_{m\beta}(x')\hatgamma_\alpha\hatgamma_\beta  +  v_{l\alpha}^*(x) v_{m\beta}(x')\hatgamma_\alpha\hatgamma^\dagger_\beta\right. \nonumber\\
&\hspace{5.5cm}\left.+ u_{l\alpha}^*(x)u_{m\beta}(x') \hatgamma_\alpha^\dagger \hatgamma_\beta + u^*_{l\alpha}(x)v_{m\beta}(x')\hatgamma_\alpha^\dagger\hatgamma_\beta^\dagger\right]( \hat{b}^\dagger+\hat{b})\nonumber\\
&= \sum_{\alpha,\beta} \biggl(\lambdatc_{\alpha\beta}\gamma^\dagger_{\alpha}\gamma_{\beta}+\lambdatd_{\alpha\beta}\gamma_{\alpha}\gamma_{\beta}\biggr)( \hat{b}^\dagger+\hat{b}) + \mathrm{H.c.},
\end{align}
where we have defined 
\begin{align}
\lambdatc_{\alpha\beta} &\equiv \frac{1}{2}\sum_{lm}\int dx \int dx'g_{lm}(x,x') [ u_{l\alpha}^*(x) u_{m\beta}(x') - v_{l\beta}^*(x) v_{m\alpha}(x')],\nonumber\\
\lambdatd_{\alpha\beta} &\equiv \sum_{lm}\int dx \int dx'g_{lm}(x,x') [v_{l\alpha}^*(x)u_{m\beta}(x')].
\end{align}

Projecting the above Hamiltonian into the lowest energy sector $\alpha = \beta = 1$ which corresponds to the ABS energy sector, we have
\begin{align}\label{eq:Heph2}
\hat{H}_{\text{e-b}} &\approx \sum_{lm}\int dx \int dx'g_{lm}(x,x') [u_{l1}^*(x)u_{m1}(x')- v_{l1}^*(x) v_{m1}(x')] \hatgamma^\dagger\hatgamma( \hat{b}^\dagger+\hat{b}) + \sum_{lm}\int dx \int dx'g_{lm}(x,x') ( \hat{b}^\dagger+\hat{b})\nonumber\\
&=\lambda \hatgamma^\dagger\hatgamma(\hat{b}^\dagger+\hat{b}) + \chi(\hat{b}^\dagger+\hat{b}),
\end{align}
where we have defined $\hatgamma \equiv \hatgamma_1$ as the Bogoliubov operator for the ABS, $\lambda \equiv 2\lambda_{11}^{(c)} $ as the ABS-boson coupling strength, and $\chi \equiv \sum_{lm}\int dx \int dx'g_{lm}(x,x') $. Note that in evaluating Eq.~\eqref{eq:Heph2}, we have used the anticommutation relation $\{\gamma,\gamma^\dagger\} = 1$, $\{\gamma,\gamma\} = 0$, and $\{\gamma^\dagger,\gamma^\dagger\} = 0$. We can eliminate the  term $\chi(\hatb^\dagger + \hatb)$ in Eq.~\eqref{eq:Heph2} by introducing the shift $\hatb \rightarrow \hatb - \chi/\Omega $ and $\hatb^\dagger \rightarrow \hatb^\dagger - \chi/\Omega$ which gives the ABS Hamiltonian as
\begin{align}
\HA &= \eabs \gamma^\dagger\gamma + \lambda \hatgamma^\dagger\hatgamma\left(\hat{b}+ \hat{b}^\dagger -2 \frac{\chi}{\Omega}\right) +  \chi\left(\hat{b}^\dagger +\hat{b}- 2\frac{\chi}{\Omega}\right) + \Omega\left( \hat{b}^\dagger -\frac{\chi}{\Omega}\right)\left( \hat{b} -\frac{\chi}{\Omega}\right)\nonumber\\
&= \left(\eabs-2 \frac{\lambda\chi}{\Omega}\right) \gamma^\dagger\gamma + \lambda \hatgamma^\dagger\hatgamma\left( \hat{b}^\dagger+\hat{b}\right) + \Omega \hat{b}^\dagger \hat{b} -\frac{\chi^2}{\Omega}. 
\end{align}
Introducing the shift $\eabs \rightarrow \eabs + 2\lambda\chi/\Omega$ and shifting the overall energy by $\frac{\chi^2}{\Omega}$, i.e., $\HA \rightarrow \HA - \frac{\chi^2}{\Omega}$, we have the Hamiltonian for the boson-coupled ABS as in Eq.~\eqref{eq:HABS} of the main text:
\begin{align}\label{eq:HABSsuppl}
\HA &= \eabs \hatgamma^\dagger \hatgamma+ \lambda \hatgamma^\dagger \hatgamma(\hatb^\dagger + \hatb)+\Omega \hatb^\dagger \hatb.
\end{align}



\section{Lang-Firsov Transformation}\label{sec:LangFirsov}
In this section, we follow Ref.~\cite{mahan2000many} to derive the matrix elements for the tunneling of electrons ($\ca^\dagger$) and holes ($\ca$) and the boson absorption or emission matrix elements $Y_{qq'}$ in Eq.~\eqref{eq:R0e} of the main text. We begin by writing the Hamiltonian of an ABS coupled to a one-dimensional normal lead and bosonic modes, e.g., phonons, plasmons, etc., as the sum of the Hamiltonian of a boson-coupled ABS, lead and tunnel coupling, $\hat{H} = \HA + \HL + \HT$ [Eq.~\eqref{eq:Hamiltonian} of the main text], where 
\begin{subequations}\label{eq:Hamiltoniansuppl}
\begin{align} 
\HA &= \eabs \hatgamma^\dagger \hatgamma+ \lambda \hatgamma^\dagger \hatgamma(\hatb^\dagger + \hatb)+\Omega \hatb^\dagger \hatb,\label{eq:HABSsuppl}\\
\HL &= \sum_{k}\varepsilon_{\mathrm{L},k} \hatclk^\dagger \hatclk, \label{eq:Hleadssuppl}\\
\HT &=  t  \hatcl^\dagger\ca + \mathrm{H.c.}\label{eq:HTsuppl}
\end{align}
\end{subequations}
Here, $\eabs $ is the ABS energy, $\hatgamma$ ($\hatgamma^\dagger$) is the Bogoliubov annihilation (creation) operator of the ABS, $\lambda$ is the ABS-boson coupling strength, $\hat{b}$ ($\hat{b}^\dagger$) is the boson annihilation (creation) operator and $\Omega$ is the boson frequency. The operator $\hatclk$ ($\hatclk^\dagger$) annihilates (creates) the lead electron with momentum $k$ and energy $\varepsilon_{\mathrm{L},k}$. The tunneling Hamiltonian $H_T$~\cite{Balatsky2006Impurity,Ruby2015Tunneling} represents the electron tunneling between the normal lead and ABS, where the electron annihilation operator of the lead and ABS at the junction given by $\hatcl = \int dk \hatclk/(2\pi)$ and $\ca$, respectively. The operator $\ca$ is obtained by projecting the operator $\hat{d}_{1}(x=0)$ [Eq.~\eqref{eq:dm}] to the ABS energy sector ($\alpha = 1$), where we have $\ca = u\gamma  + v\gamma^\dagger$. For notational simplicity, here we define $\gamma \equiv \gamma_1$, $u \equiv u_{\mathrm{11}}(x=0)$ and $v \equiv v_{\mathrm{11}}(x=0)$ where $u$ and $v$ are the particle and hole components of the ABS wave function at the junction ($x = 0$). In this paper, we renormalize the ABS wave function such that $|u|^2+ |v|^2 = 1$. Note that since we consider only the subgap state and ignore the above-gap states, the relation $\hat{d}_{\mathrm{A}} = u\gamma + v\gamma^\dagger$ is  only approximate which makes $\hat{d}_{\mathrm{A}}$ nonfermionic. The operator $\ca$ becomes fermionic if all the states in the superconductor including the above-gap states are taken into account [see Eq.~\eqref{eq:dm}]. Our conclusion on the PHS breaking of the subgap conductance due to the ABS-boson coupling does not rely on the fermionic properties of $\ca$.

To eliminate the ABS-boson coupling, we can transform the Hamiltonian~[Eq.~\eqref{eq:Hamiltoniansuppl}]  using a canonical transformation 
\begin{equation}
\hat{\tilde{H}} = e^{\hatS} \hatH e^{-\hatS}, 
\end{equation}
where 
\begin{align}
\hatS&= \frac{\lambda}{\Omega} \hatgamma^\dagger \hatgamma (\hatb^\dagger - \hatb)
\end{align}
is the Lang-Firsov transformation operator~\cite{lang1963zh}. Using the relation
\begin{align}
\hat{\tilde{A}} &= e^{\hatS} \hatA e^{-\hatS} = \hatA + [\hatS,\hatA] + \frac{1}{2!}[\hatS,[\hatS,\hatA]]+\cdots,
\end{align}
we can write the transformed annihilation and creation operators for the Bogoliubov quasiparticles, electrons and bosonic modes as
\begin{subequations}\label{eq:cantrans}
\begin{align}
\hattildegamma &= \hatgamma \hatY,\\
\hattildegamma^\dagger &= \hatgamma^\dagger \hatY^\dagger,\\
\cat &= u\hatgamma \hatY + v\hatgamma^\dagger \hatY^\dagger, \\
\cat^\dagger&= u^*\hatgamma^\dagger \hatY^\dagger +v^*\hatgamma \hatY,\\
\hattildeb &= \hatb - \frac{\lambda}{\Omega}\hatgamma^\dagger \hatgamma,\\
\hattildeb^\dagger &= \hatb^\dagger - \frac{\lambda}{\Omega}\hatgamma^\dagger \hatgamma,
\end{align}
\end{subequations}
where $\hatY = \mathrm{exp}\left[-\frac{\lambda}{\Omega}(\hatb^\dagger - \hatb)\right]$. Under this transformation, the number operator remains the same, i.e., $\hattildegamma^\dagger \hattildegamma = \hatgamma^\dagger \hatgamma \hat{Y}^\dagger \hat{Y} = \hatgamma^\dagger \hatgamma$ and the Hamiltonians~[Eq.~\eqref{eq:HABSsuppl} and Eq.~\eqref{eq:HTsuppl}] transform as
\begin{subequations}
\begin{align}
\HAtilde &= \eabs \hatgamma^\dagger \hatgamma + \lambda\left(\hatb^\dagger + \hatb - \frac{2\lambda}{\Omega}\hatgamma^\dagger \hatgamma\right)\hatgamma^\dagger \hatgamma + \Omega\left(\hatb^\dagger - \frac{\lambda}{\Omega}\hatgamma^\dagger \hatgamma \right)\left(\hatb - \frac{\lambda}{\Omega}\hatgamma^\dagger \hatgamma \right)\\
&= \left(\eabs -\frac{\lambda^2}{\Omega}\right)\hatgamma^\dagger \hatgamma + \Omega \hatb^\dagger \hatb,\nonumber\\
\HTtilde &=   t \hatcl^\dagger \cat+\mathrm{H.c.},\label{eq:HTdirect}
\end{align}
\end{subequations}
where $\cat = u\hatgamma\hatY + v\hatgamma^\dagger\hatY^\dagger$ is the Lang-Firsov transformation of $\ca$. 

We can evaluate the matrix elements for the electron and hole tunneling which change the ABS occupancy number $n$ from $0 \rightarrow 1$ and the boson occupancy from $q \rightarrow q'$ as
\begin{subequations}
\begin{align}
\langle 1,q'|\cat^\dagger|0,q\rangle &= u^*\langle 1|\hatgamma^\dagger |0\rangle   \langle q'|\hatY^\dagger|q\rangle= u^*Y_{qq'}, \\
\langle 1,q'|\cat|0,q\rangle &= v\langle 1|\hatgamma^\dagger|0\rangle  \langle q'|\hatY^\dagger|q\rangle  = v Y_{qq'},
\end{align}
\end{subequations}
respectively. Using the Baker-Campbell-Haussdorf formula, we have
\begin{equation}
\hatY^\dagger = e^{\frac{\lambda}{\Omega}(\hatb^\dagger - \hatb)}= e^{-\frac{\lambda^2}{2\Omega^2}}e^{\frac{\lambda}{\Omega} \hatb^\dagger}e^{-\frac{\lambda}{\Omega} \hatb},
\end{equation}
and we can evaluate the boson emission or absorption matrix element as~\cite{Mitra2004Phonon}
\begin{align}\label{eq:Yqqp}
Y_{qq'} \equiv \langle q'|\hatY^\dagger|q \rangle = \langle q'|e^{\frac{\lambda}{\Omega}(b^\dagger - b)}|q \rangle &= \langle q'|  e^{-\frac{\lambda^2}{2\Omega^2}}e^{\frac{\lambda}{\Omega} \hatb^\dagger}e^{-\frac{\lambda}{\Omega} \hatb} |q\rangle\nonumber\\
 &=  e^{-\frac{\lambda^2}{2\Omega^2}} \sum_{m=0}^{\mathrm{min}(q,q')}\left(\frac{\lambda}{\Omega}\right)^{q'-m}\left(-\frac{\lambda}{\Omega}\right)^{q-m}\frac{\sqrt{q!q'!}}{m!(q-m)!(q'-m)!},
\end{align}
where $|Y_{qq'}|^2$ is symmetric under the interchange $q \leftrightarrow q'$. 
Note that in going to the second line of Eq.~\eqref{eq:Yqqp}, we have used the following relations:
\begin{subequations}\label{eq:Yqqeq}
\begin{align}
e^{-\frac{\lambda}{\Omega} \hatb}|q\rangle &= \sum_{m = 0}^\infty \frac{1}{m!} \left(-\frac{\lambda}{\Omega}\right)^m \hatb^m|q\rangle =\sum_{m =0}^{q} \frac{1}{m!}\left(-\frac{\lambda}{\Omega}\right)^m\sqrt{\frac{q!}{(q-m)!}} |q-m\rangle,\\
\langle q'|e^{\frac{\lambda}{\Omega} \hatb^\dagger} &= \sum_{l = 0}^{\infty} \frac{1}{l!}\left(\frac{\lambda}{\Omega}\right)^l \langle q'|(\hatb^\dagger)^l = \sum_{l = 0}^{q'} \frac{1}{l!}\left(\frac{\lambda}{\Omega}\right)^l\sqrt{\frac{q'!}{(q'-l)!}}\langle q'-l|.
\end{align}
\end{subequations}

\section{Rate equation and tunneling rates}

The stationary-state rate equation satisfied by the probability $\Pnq$ of an ABS-boson system being in the state $|n,q\rangle$, i.e., having an ABS occupation number $n$ and boson occupation number $q$, is given by~\cite{Mitra2004Phonon,Koch2004Thermo}
\begin{align}\label{eq:dP1q}
0 &= \frac{\partial\Pnq}{\partial t}\nonumber\\
&=\sum_{q'} \Pnpqp\left[ \Rnpe + \Rnph\right] -\Pnq \sum_{q'}\left[\Rne+ \Rnh\right] \nonumber\\
&\hspace{1cm}+ P^{n}_{q+1}\eta_{q+1;-} +  P^{n}_{q-1}\eta_{q-1;+}-\Pnq\left(\eta_{q;+}+\eta_{q;-}\right).
\end{align} 
The second line in Eq.~\eqref{eq:dP1q} represents the probability flux due to hopping of an electron ($e$) or hole ($h$) from the lead to the ABS which changes the ABS occupation number from $\bar{n} \equiv 1- n$ to $n$ and the boson occupancy from $q'$ to $q$ and vice versa. The quantity $P_q^n$ denotes the probability that the system is in the state $|n,q\rangle$ and $R_{q\rightarrow q'}^{n\rightarrow\bar{n}}$ denotes the transition rate from the state $|n,q\rangle$ to the state $|\bar{n},q'\rangle$. The third line of Eq.~\eqref{eq:dP1q} represents the boson relaxation where the boson emission and absorption probabilities are $\eta_{q;+} = A(q+1) $ and $\eta_{q;-}= Bq$, respectively, with $A = Be^{-\Omega/\kb T}$. These probability rates are consistent with the fluctuation-dissipation theorem. If the boson relaxation rate is faster than the tunneling rate $\Gamma/\hbar$ such that the bosons acquire the equilibrium distribution $\Pphq = e^{-q\Omega/\kb T}(1-e^{-\Omega/\kb T})$, the probability $\Pnq$ can be factorized as $\Pnq=\Pn\Pphq$. 
Summing Eq.~\eqref{eq:dP1q} over $q$ for these factorized probabilities gives
\begin{align}\label{eq:dP1tot}
0 = \Pnb ( \Rtnbe + \Rtnbh) - P^{n} \left(\Rtne+ \Rtnh\right)+\Pn\sum_{q}\left[P^{\mathrm{b}}_{q+1}\eta_{q+1;-} +  P^{\mathrm{b}}_{q-1}\eta_{q-1;+} - P^{\mathrm{b}}_{q} \left(\eta_{q;+}+\eta_{q;-} \right)\right],
\end{align} 
where $R^{n \rightarrow \bar{n};e(h)} \equiv \sum_{q,q'}\Pphq R^{n \rightarrow \bar{n};e(h)}_{q\rightarrow q'}$ and $R^{\bar{n} \rightarrow n;e(h)} \equiv \sum_{q,q'}\Pphqp R^{\bar{n} \rightarrow n;e(h)}_{q'\rightarrow q}$.
Since the sum of the boson relaxation terms over $q$ [the last four terms in Eq.~\eqref{eq:dP1tot}] is zero, Eq.~\eqref{eq:dP1tot} then reduces to Eq.~\eqref{eq:dP1} of the main text. 

For the tunneling Hamiltonian in Eq.~\eqref{eq:HTdirect}, the rates of the electron and hole tunneling processes can be calculated from Fermi's Golden Rule to be
\begin{subequations}\label{eq:R0esuppl}
\begin{align}
\Rne &= \frac{2 \pi t^2\nu_0}{\hbar}   \left|\left\langle \bar{n},q'\left| \cat^\dagger\right|n,q\right\rangle\right|^2 f(E_{\bar{n},q'} - E_{n,q} - eV)\nonumber\\
&=\frac{\Gamma}{\hbar} |\langle \bar{n}| \ca^\dagger|n\rangle|^2|Y_{qq'}|^2 f(E_{\bar{n},q'} - E_{n,q} - eV),\\
\Rnh &= \frac{2 \pi t^2 \nu_0}{\hbar} \left|\left\langle \bar{n},q'\left| \cat \right|n,q\right\rangle\right|^2 f(E_{\bar{n},q'} - E_{n,q} + eV)\nonumber\\
&=\frac{\Gamma}{\hbar}|\langle \bar{n}| \ca|n\rangle|^2 |Y_{qq'}|^2 f(E_{\bar{n},q'} - E_{n,q} + eV),
\end{align}
\end{subequations}
where $\langle \bar{n}|\ca^\dagger |n \rangle$ and $\langle \bar{n}|\ca |n \rangle$ are the bare tunneling matrix elements for electrons and holes,  respectively, $Y_{qq'} = \langle q'|e^{-\lambda(\hatb^\dagger-\hatb)/\Omega}|q\rangle$ is the boson emission or absorption matrix element, and $f(E) = [1+\mathrm{exp}({E/k_{\mathrm{B}}T})]^{-1}$ is the lead Fermi function.

\section{Details on Model I. Tunneling into Boson-Coupled ABS }
\subsection{Proof for the particle-hole asymmetry of boson-coupled-ABS conductance}\label{sec:proof}
In this section, we will prove that, unless $|u| = |v|$, the current into a boson-coupled ABS [Eq.~\eqref{eq:currentrate1} of the main text] is in general \textit{not} PH antisymmetric, i.e., $I(V_0) \neq I(-V_0)$ resulting in a PH \textit{asymmetric} conductance, i.e., $\left.\frac{dI}{dV}\right|_{V = V_0} \neq \left.\frac{dI}{dV}\right|_{V = -V_0}$. Substituting the transition rates
\begin{subequations}
\begin{align}
\Rtzeroe &= \Gamma|u|^2 W(\eabsmtilde)/\hbar,\\
\Rtonee &= \Gamma|v|^2  W(-\eabsptilde)/\hbar,\\
\Rtzeroh &= \Gamma|v|^2 W (\eabsptilde)/\hbar,\\
\Rtoneh &= \Gamma|u|^2  W(-\eabsmtilde)/\hbar,
\end{align}
\end{subequations}
into Eq.~\eqref{eq:R0e} of the main text, we can evaluate the current [Eq.~\eqref{eq:currentrate1} of the main text] as
\begin{align}\label{eq:IW}
I &= \frac{2e}{\hbar}  \frac{\Gamma|uv|^2\left[W(\eabsmtilde) W(-\eabsptilde) - W (\eabsptilde)W(-\eabsmtilde)\right]}{ |u|^2 F(\eabsmtilde) + |v|^2  F(\eabsptilde)},
\end{align}
where
\begin{subequations}
\begin{align}
W(x) &=  \sum_{q,q'}\Pphq |Y_{qq'}|^2  f(x- \Omega(q -q'))\label{eq:wx},\\
\eabspmtilde &= \eabs - \lambda^2/\Omega \pm eV,\\
F(x) &= W(x) + W(-x),\label{eq:Vx}
\end{align}
\end{subequations} 
with $f(x) = [1+\mathrm{exp}({x/k_{\mathrm{B}}T})]^{-1}$ being the Fermi function.

We will prove below that the function $F(x)$ in the denominator of Eq.~\eqref{eq:IW} is an increasing function of $x$ and hence the denominator in Eq.~\eqref{eq:IW} is asymmetric with respect to the interchange $V \leftrightarrow -V$ unless $|u| = |v|$. By rewriting $W(x)$ in Eq.~\eqref{eq:wx} as
\begin{align}
W(x) &=  \int_{-\infty}^{\infty} d\omega Q(\omega) f(x-\omega),
\end{align}
where
\begin{equation}
Q(\omega) = \sum_{q,q'}\Pphq |Y_{qq'}|^2 \delta(\omega - \Omega(q -q') ),
\end{equation}
we have
\begin{align}\label{eq:Vx}
F(x) \equiv W(x) + W(-x) &= \int_{-\infty}^{\infty}  [Q(\omega) - Q(-\omega)]f(x-\omega) d\omega  + \sum_{q,q'}\Pphq|Y_{qq'}|^2 \nonumber\\
&= \int_0^\infty [Q(\omega) - Q(-\omega)][f(x-\omega) - f(x+\omega)] d\omega + \sum_{q,q'}\Pphq|Y_{qq'}|^2 \nonumber\\
&= \frac{1}{2}\int_0^\infty \left[Q(\omega) - Q(-\omega)\right]\left[\mathrm{tanh}\left(\frac{x+\omega}{2\kb T}\right) - \mathrm{tanh}\left(\frac{x-\omega}{2 \kb T}\right)\right]d\omega +  \sum_{q,q'}\Pphq|Y_{qq'}|^2.
\end{align}
In the following, we will prove that $F(x)$ is a monotonic function of $x$. We first begin by noting that $Q(\omega) - Q(-\omega) \leq 0$ for $\omega \geq 0$. The proof is as follows
\begin{align}\label{eq:Zomega}
Q(\omega) - Q(-\omega) &= \sum_{q,q'} \Pphq |Y_{qq'}|^2\left[\delta(\omega - \Omega(q-q')) - \delta(-\omega - \Omega(q-q'))\right]\nonumber\\
&= \sum_{q,q'} \Pphq |Y_{qq'}|^2\left[\delta(\omega - \Omega(q-q')) - \delta(\omega +\Omega(q-q'))\right]\nonumber\\
&= \sum_{q,q'} (\Pphq - \Pphqp) |Y_{qq'}|^2\delta(\omega - \Omega(q-q')),
\end{align}
where in the third line we interchange $q$ with $q'$ for the second sum and use $|Y_{qq'}|^2 = |Y_{q'q}|^2$. For $\omega \geq 0$, the delta function forces $q \geq q'$ implying that $(\Pphq - \Pphqp) \propto \mathrm{exp}(-q\Omega/\kb T) - \mathrm{exp}(-q'\Omega/\kb T) \leq 0$. As a result, $Q(\omega) - Q(-\omega) \leq 0$.

To prove that $F(x)$ is a monotonic function of $x$, we take the derivative of $F(x)$ [Eq.~\eqref{eq:Vx}] with $x$ which gives
\begin{align}
F'(x) = \frac{1}{4\kb T}\int_0^\infty \left[Q(\omega) - Q(-\omega)\right]\left[\mathrm{sech}^2\left(\frac{x+\omega}{2\kb T}\right) - \mathrm{sech}^2\left(\frac{x-\omega}{2\kb T}\right)\right]d\omega \geq 0.
\end{align}
So, $F(x)$ increases monotonically with $x$. This means that unless $|u| = |v|$, the denominator in Eq.~\eqref{eq:IW} is PH asymmetric with respect to the interchange of $V \leftrightarrow -V$ [which amounts to interchanging $\eabsmtilde \leftrightarrow \eabsptilde$ in Eq.~\eqref{eq:IW}]. To prove that the boson-assisted tunneling model (model II) can also break the PHS of subgap conductances, we simply replace $|Y_{qq'}|$ by $|X_{qq'} - \lambda Y_{qq'}/\Omega|$ in the above derivation, where $X_{qq'} \equiv e^{-\frac{\lambda^2}{2\Omega^2}} \langle q'| e^{\frac{\lambda}{\Omega} \hatb^\dagger}(\hatb^\dagger + \hatb) e^{-\frac{\lambda}{\Omega} \hatb}|q\rangle$ [Eq.~\eqref{eq:Xqq}]. Even though the conductance is not PH symmetric, under a simultaneous interchange of $V \leftrightarrow -V$ and $|u| \leftrightarrow |v|$, the current is antisymmetric ($I \rightarrow -I$) resulting in a symmetric conductance. This means that the conductance for $|v|^2 > |u|^2$ can be obtained from the conductance for $|v|^2 < |u|^2$ (shown in Figs.~\ref{fig:currentrate} and~\ref{fig:Keldyshcurrent} of the main text) by interchanging both $|u|\leftrightarrow |v|$ and $V \leftrightarrow -V$ simultaneously. As a result, the higher and lower peaks switch sides when $|v| \leftrightarrow |u|$, which changes the sign of the PH asymmetry $\zeta$.

The PH asymmetry of the conductance can be understood more intuitively in the limit of large positive and negative voltages $|eV| \gtrsim |\eabstilde| + \kb T$. In the large-positive-voltage regime ($eV \gtrsim |\eabstilde| + \kb T$), hole tunneling processes are energetically forbidden  [$\Rtzeroh, \Rtoneh \approx 0$ since $W (\eabsptilde),W(-\eabsmtilde) \approx 0$]. On the other hand, in the large-negative-voltage regime where $eV \lesssim -(|\eabstilde| +\kb T)$, electron tunneling processes are not energetically allowed [$\Rtzeroe, \Rtonee \approx 0$ since $W(\eabsmtilde), W(-\eabsptilde) \approx 0$]. In this limit, the current [Eq.~\eqref{eq:IW}] thus reduces to Eq.~\eqref{eq:currentrate2} of the main text:
\begin{align}\label{eq:Imain}
I = \begin{cases}
2e \dfrac{\Rtzeroe \Rtonee}{\Rtzeroe + \Rtonee} & \text{ for $eV \gtrsim |\eabstilde| + \kb T$,}\\[1em]
 -2e \dfrac{\Rtzeroh \Rtoneh}{\Rtzeroh + \Rtoneh} & \text{ for $eV \lesssim -(|\eabstilde| +\kb T)$.}
\end{cases}
\end{align} 
implying that the current at large positive and negative voltages are due to sequential tunnelings of electrons and holes, respectively (see Fig.~\ref{fig:schematic} of the main text). For both the boson-coupled ABS model and boson-assisted tunneling into ABS model, the current is in general not PH antisymmetric, i.e., $I(-V_0) \neq I(V_0)$ or the conductance is PH asymmetric ($\left.\frac{dI}{dV}\right|_{V = V_0} \neq \left.\frac{dI}{dV}\right|_{V = -V_0}$) because of the rate asymmetry between the first and second tunneling processes of electrons and holes (i.e., $\Rtzeroe \neq \Rtoneh$ and $\Rtonee \neq \Rtzeroh$). This rate asymmetry arises because the second tunneling process which happens at energy deep inside the Fermi level is energetically allowed to emit more bosons hence occurs with a larger rate than the first tunneling process. Without the ABS-boson coupling ($\lambda = 0$), $W(x)  = f(x)$ where $f(x)$ is the Fermi function and the current [Eq.~\eqref{eq:IW}] is $I = 2\frac{e}{\hbar}\Gamma |uv|^2 [f(\eabsm)-f(\eabsp)]$ which is PH antisymmetric, i.e., $I(V) = -I(-V)$. Thus, the conductance into ABSs in gapped superconductors is PH symmetric.
\subsection{Proof for the temperature independence of the conductance peak area}\label{sec:proofarea}
While the conductance for boson-coupled ABSs is in general PH asymmetric, the conductance peak areas calculated using the boson-coupled ABS model in Sec.~\ref{sec:direct} of the main text are independent of temperature and equal for both negative and positive voltages. To see this, we can calculate the current at $ V = \pm \infty$ by using Eq.~\eqref{eq:Imain}. Note that for $eV \gg |\eabstilde| + \kb T$, we have $f(\eabsmtilde) = f(-\eabsptilde) = 1$ and for $eV \ll |\eabstilde| + \kb T$, $f(\eabsptilde) = f(-\eabsmtilde) = 1$. This in turn yields $W(\eabsmtilde) = W(-\eabsptilde) =W(\eabsptilde) = W(-\eabsmtilde) = \sum_{q,q'}\Pphq |Y_{qq'}|^2 = 1$. Using $|u|^2+|v|^2 = 1$, we then have the current [Eq.~\eqref{eq:Imain}] as
\begin{align}
I = \begin{cases}
\dfrac{e}{2\hbar} \Gamma [1- (|u|^2-|v|^2)^2]  & \text{ for $eV \gg |\eabstilde| + \kb T$,}\\[1em]
 -\dfrac{e}{2\hbar} \Gamma [1- (|u|^2-|v|^2)^2] & \text{ for $eV \ll -(|\eabstilde| +\kb T)$.}
\end{cases}
\end{align} 
So, $I(V = \infty) = -I(V = -\infty)$ and both $I(V = \pm\infty)$ are independent of temperature. Since the current magnitude at large voltages $|I(V = \pm\infty)|$ is the area under the conductance peak, this means that the conductance peak area for positive and negative voltages are equal and independent of temperature. This fact can also be seen from the current plots in Figs.~\ref{fig:currentrateappendix}(c) and ~\ref{fig:currentKeldyshappendix}(c) which are calculated using the rate equation and Keldysh approach, respectively. Contrary to model I, the boson-assisted-tunneling model (model II) gives rise to temperature-dependent conductance peak area (see Sec.~\ref{sec:indirect} of the main text).

\section{Current calculated from the rate equation and Keldysh approach}
\begin{figure}[h!]
\centering
\includegraphics[width=0.6\linewidth]{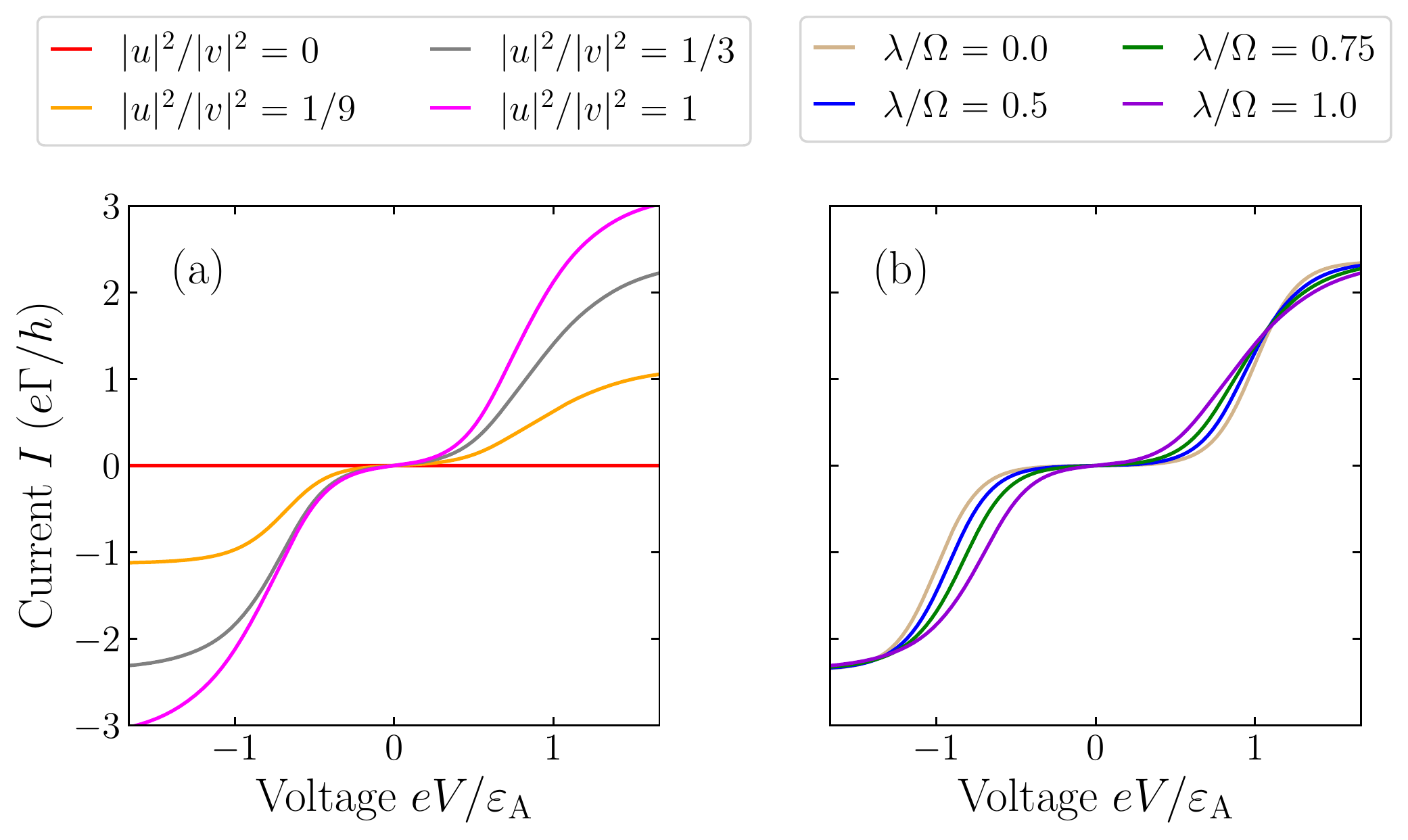}
\caption{Current $I$ of boson-coupled ABSs vs voltage $V$ calculated using the rate equation [Eq.~\eqref{eq:currentrate1} of the main text] for (a) different ratios of PH components $|u|^2/|v|^2$ with $\lambda/\Omega = 1$ and (b) different ABS-boson coupling strengths $\lambda$ with $|u|^2/|v|^2 = 1/3$. The conductance calculated from the above current is shown in Fig.~\ref{fig:currentrate} of the main text. The parameters for all panels are: $\eabs/\Omega = 3$, $\Gamma/\Omega = 0.05/(2\pi)$, and $\kb T/\Omega = 0.4$.}\label{fig:currentonlyrate}
\end{figure}

\begin{figure}[h!]
\centering
\includegraphics[width=0.6\linewidth]{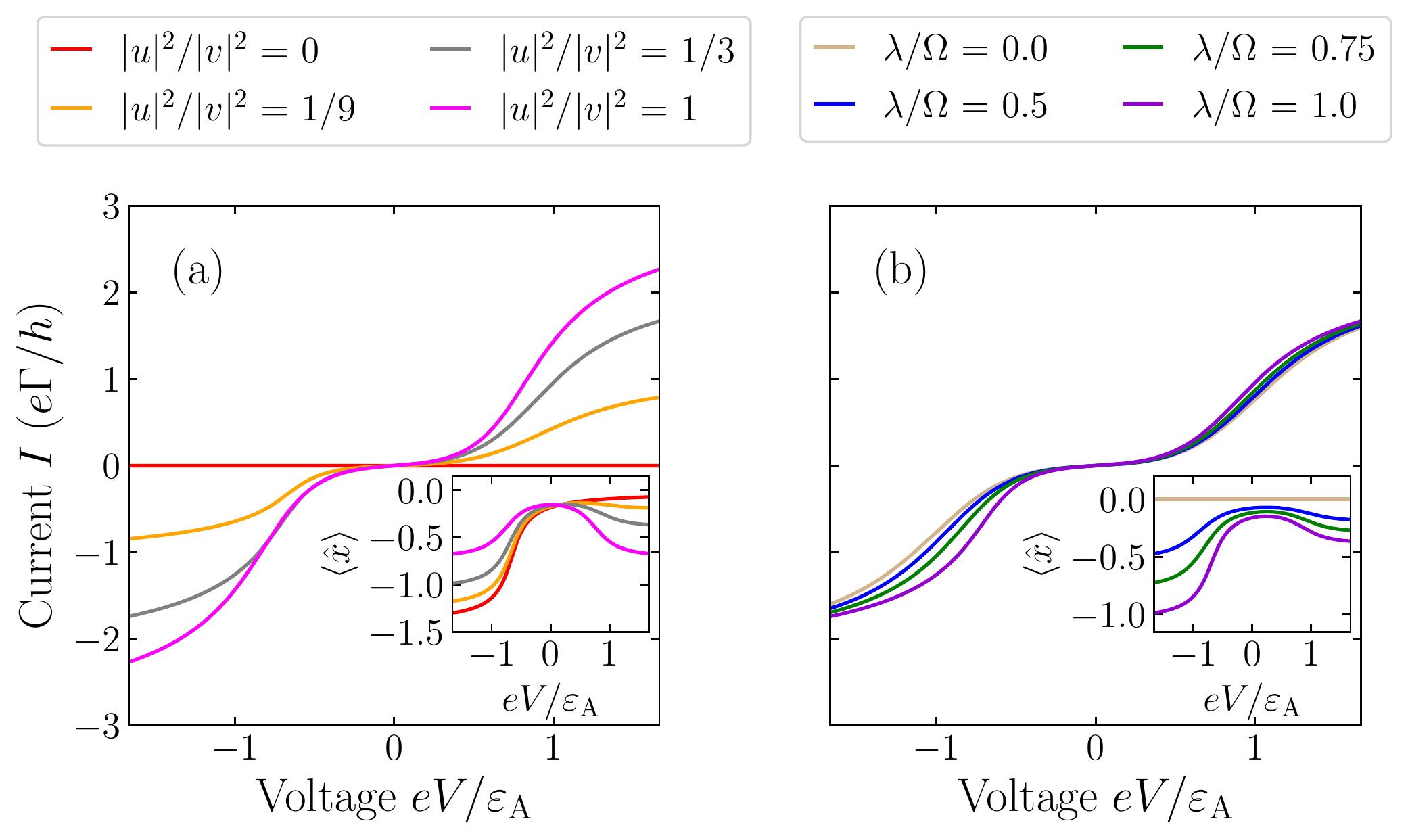}
\caption{Current $I$ of boson-coupled ABSs vs voltage $V$ calculated using the mean-field Keldysh approach [Eq.~\eqref{eq:IV} of the main text] for (a) different ratios of PH components $|u|^2/|v|^2$ with $\lambda/\Omega = 1$ and (b) different ABS-boson coupling strengths $\lambda$ with $|u|^2/|v|^2=1/3$. The conductance calculated from the above current is shown in Fig.~\ref{fig:Keldyshcurrent} of the main text. Inset: Expectation value of the boson displacement operator $\langle \hat{x} \rangle$ vs voltage $V$ calculated self-consistently using Eq.~\eqref{eq:selfconst} of the main text. The parameters for all panels are: $\eabs/\Omega = 3$, $\Gamma/\Omega = 2$, and $\kb T/\Omega = 0.4$.}\label{fig:currentonlyKeldysh}
\end{figure}

In this section, we show the current calculated from the rate equation (Fig.~\ref{fig:currentonlyrate}) and mean-field Keldysh approach~(Fig.~\ref{fig:currentonlyKeldysh}) corresponding to the conductance shown in Figs.~\ref{fig:currentrate} and~\ref{fig:Keldyshcurrent} of the main text, respectively.
As shown in Figs.~\ref{fig:currentonlyrate}(a) and ~\ref{fig:currentonlyKeldysh}(a), the current decreases with increasing ABS's PH content imbalance $||u|^2 - |v|^2|$ where $I = 0$ when $||u|^2 - |v|^2| = 1$. This is due to the fact that the terms $\Rtzeroe \Rtonee$ and $\Rtzeroh \Rtoneh$ in the current expression [Eq.~\eqref{eq:currentrate1} of the main text] are $\propto |uv|^2 = [1 - (|u|^2 - |v|^2)^2]/4$. 

\section{Dependence of the current and conductance calculated from the rate equation on ABS-boson coupling strength, temperature, boson frequency, and ABS energy}

\begin{figure}[h!]
\centering
\includegraphics[width=\linewidth]{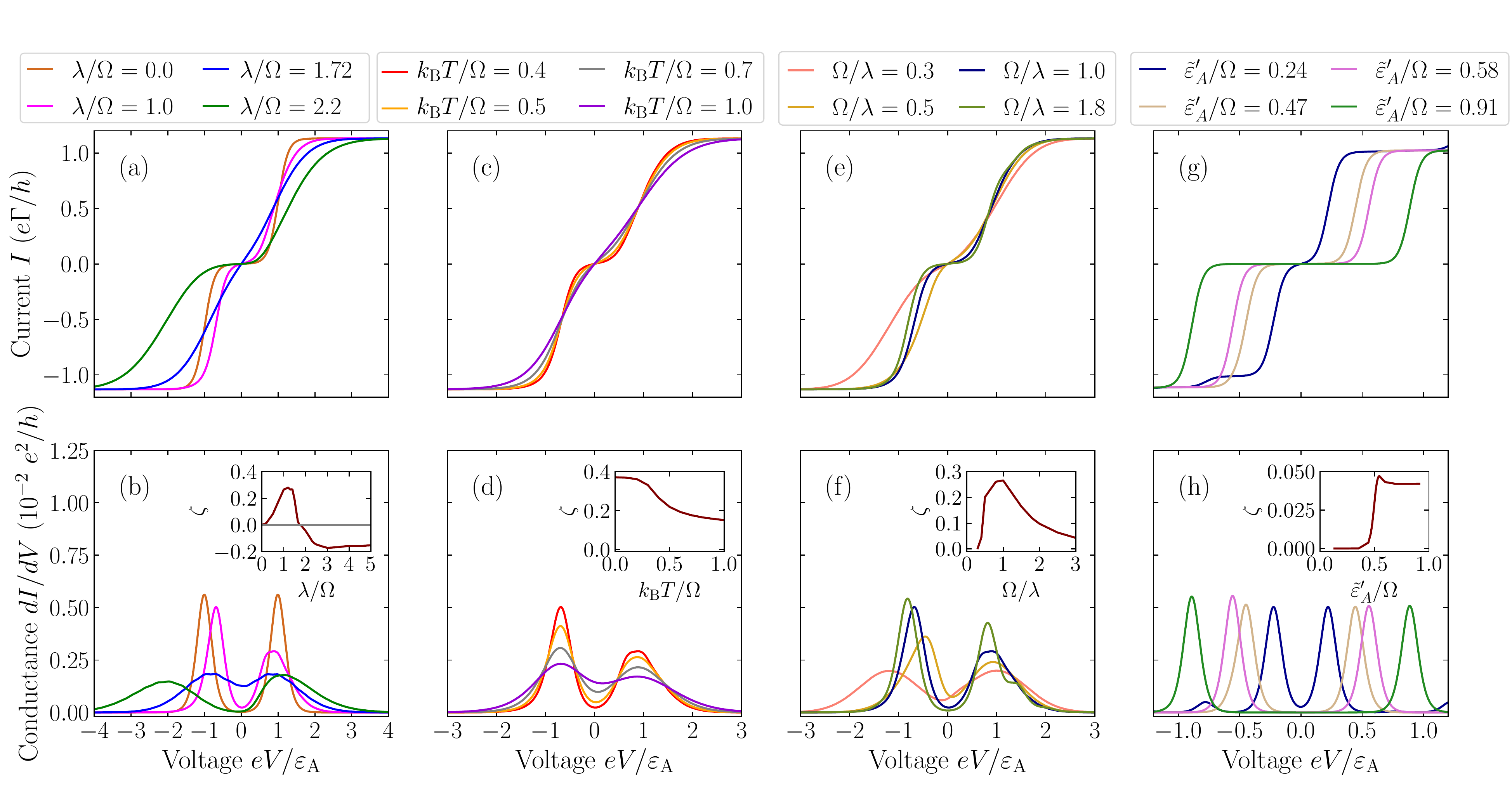}
\caption{Current $I$ (Upper panels) and conductance $dI/dV$ (Lower panels) of boson-coupled ABSs vs voltage $V$ calculated using the rate equation [Eq.~\eqref{eq:currentrate1} of the main text] for (a,b) different ABS-boson coupling strengths $\lambda$ with $\kb T = 0.4$, $\eabs = 3$, $\Gamma = 0.05/(2\pi)$, and $\Omega = 1.0$, (c,d) different temperatures $T$ with $\lambda = 1.0$, $\eabs = 3.0$, $\Gamma = 0.05/(2\pi)$, and $\Omega = 1.0$, (e,f) different boson frequencies $\Omega$ with  $\kb T = 0.4$, $\eabs = 3.0$, $\Gamma = 0.05/(2\pi)$, and $\lambda = 1.0$, and (g,h) different renormalized ABS energies $\eabstilde' \equiv |\eabs - \lambda^2/\Omega| + \kb T/2$ with $\kb T = 0.4$, $\lambda=3.0$ and $\Gamma = 1/(40\pi)$, and $\Omega = 9.0$. Inset: (b) Conductance PH asymmetry $\zeta$ vs $\lambda/\Omega$, (d) $\zeta$ vs temperature $T$,  (f) $\zeta$ vs boson frequency $\Omega$, and (h) $\zeta$ vs $\eabstilde'/\Omega$. Panel (h) shows that the subgap conductances exhibit PH asymmetry only for $\eabstilde'/\Omega \gtrsim 0.5$ where $\eabstilde' \equiv |\eabs - \lambda^2/\Omega| + \kb T/2$. The parameters used for all panels are: $|u|^2/|v|^2 = 1/9$.}\label{fig:currentrateappendix}
\end{figure}

Figure~\ref{fig:currentrateappendix} shows the current (upper panels) and conductance (lower panels) of boson-coupled ABSs calculated from the rate equation~[Eq.~\eqref{eq:currentrate1} of the main text] for different ABS-boson coupling strengths $\lambda$ [Figs.~\ref{fig:currentrateappendix}(a,b)], temperatures $T$ [Figs.~\ref{fig:currentrateappendix}(c,d)], boson frequencies $\Omega$ [Figs.~\ref{fig:currentrateappendix}(e,f)] and ABS energies $\eabs$ [Figs.~\ref{fig:currentrateappendix}(g,h)]. Figure~\ref{fig:currentrateappendix}(b) shows that the magnitude of the conductance PH asymmetry $\zeta$ has a nonmonotonic dependence on the ABS-boson coupling strength $\lambda$. In the limit $\eabs -\lambda^2/\Omega \gg \kb T$ (where the two conductance peaks are well separated), the conductance PH asymmetry $\zeta$ increases with increasing $\lambda$; this corresponds to the results shown in Fig.~\ref{fig:currentrate}(b) of the main text. As $\lambda$ keeps increasing, the two conductance peaks approach each other and in the regime where $\eabs -\lambda^2/\Omega < \kb T$, the two peaks start to overlap with each other and $\zeta$ decreases with increasing $\lambda$. Note that in the regime where $\eabs -\lambda^2/\Omega > 0$, the higher peak is at positive voltage for the case where $|u|^2 > |v|^2$ while for the case where $|v|^2 > |u|^2$, the higher peak is at negative voltage. When $\eabs -\lambda^2/\Omega = 0$, the two conductance peaks merge at the zero voltage which gives a zero conductance PH asymmetry ($\zeta = 0$). Increasing $\lambda$ beyond this point splits the peaks but with the low and high peaks now switching sides which in turn changes the sign of $\zeta$. As $\lambda$ increases further, the two peaks move away from each other and the PH asymmetry $\zeta$ increases in magnitude; beyond a certain value of $\lambda$, $\zeta$ becomes weakly dependent on $\lambda$ as shown in the inset of Fig.~\ref{fig:currentrateappendix}(b). Note that for large enough $\lambda$, the position of the conductance peaks are no longer PH symmetric [see green curve in Fig.~\ref{fig:currentrateappendix}(b)].


Figure~\ref{fig:currentrateappendix}(d) shows that the conductance PH asymmetry $\zeta$ decreases with increasing temperature $T$. This is due to the fact that temperature broadens the conductance peaks.  The dependence of the ABS conductance on the boson frequency $\Omega$ is shown in Fig.~\ref{fig:currentrateappendix}(f). The PH asymmetry $\zeta$ has a nonmonotonic behavior with the boson frequency $\Omega$ where it first increases with increasing $\Omega$ and then after reaching its maximum, it decreases with increasing $\Omega$. The initial increase of $\zeta$ with increasing $\Omega$ can be attributed to the fact that the two conductance peaks move away from each other as $\Omega$ increases ($\eabstilde = \eabs - \lambda^2/\Omega$ increases with increasing $\Omega$). The decrease of $\zeta$ for large $\Omega$ is due to the fact that the effective ABS-boson coupling strength $\lambda/\Omega$ decreases with increasing $\Omega$. Figure~\ref{fig:currentrateappendix}(h) shows the dependence of the conductance on the ABS energy. As shown in the inset of panel (h), the peak conductance only exhibits the PH asymmetry for $\Omega \lesssim  2|\eabstilde| + \kb T$. This can be understood from the fact that the second tunneling process [whose rate is $\Rtonee$ in Fig.~\ref{fig:schematic}(a) or $\Rtoneh$ in Fig.~\ref{fig:schematic}(b) of the main text] can transfer lead electrons or holes with an energy difference up to $\sim 2 |\eabstilde| + \kb T$ from the subgap state, where this energy difference is transferred in form of boson energy $\Omega$. Even though in this paper, we focus only on the regime $\kb T \gtrsim \lambda$ where the boson sidebands vanish due to the thermal broadening~\cite{Mitra2004Phonon}, the dependence of the ABS conductance peak on the above parameters also hold true in the case where there are boson sidebands. Moreover, the PH asymmetry of the boson sidebands also have similar dependences on the above parameters as that of the ABS conductance peak.

\section{Derivation of the current in the Keldysh formalism}
In this section, we derive the current [Eq.~\eqref{eq:IV} of the main text] following Refs.~\cite{Cuevas1996Hamiltonian,Gonzalez2020Photon,Ruby2015Tunneling}. We begin by writing the Hamiltonian as 
\begin{equation}\label{eq:Htotal}
\hat{H} =  \HA + \HL +\HT, 
\end{equation}
where
\begin{subequations}
\begin{align}
\HA &= (\eabs + \sqrt{2}\lambda \avex) \hatgamma^\dagger \hatgamma,\label{eq:HAmean}\\
\HL &= \sum_{k}\varepsilon_{\mathrm{L},k} \hatclk^\dagger \hatclk, \\
\HT &=  t  \hatcl^\dagger\ca + \mathrm{H.c.}
\end{align}
\end{subequations}
In Eq.~\eqref{eq:HAmean}, $\HA$ is the mean-field Hamiltonian of the ABS-boson system  obtained by replacing $\hatx$ in Eq.~\eqref{eq:HABSa} of the main text by $\avex$ where $\avex = \frac{ \langle\hatb + \hatb^\dagger\rangle}{\sqrt{2}}$ is the mean-field boson displacement. Note that we have dropped the constant in the Hamiltonian $\HA$ in Eq.~\eqref{eq:HAmean} since this is just a shift in the energy. To calculate the current, we first apply a gauge transformation~\cite{rogovin1974fluctuation} 
\begin{equation}
\hat{U}(\tau) = \mathrm{exp}\left\{\frac{i}{\hbar} \int_0^\tau d\tau' [\mu_\mathrm{L}(\tau')\hat{N}_\mathrm{L} + \mu_\mathrm{S}(\tau')\hat{N}_\mathrm{S}] \right\},
\end{equation}
to the Hamiltonian $\hat{H}$ in Eq.~\eqref{eq:Htotal}, where $\hat{N}_\mathrm{L} = \hatcl^\dagger \hatcl$ and $N_\mathrm{S} = \ca^\dagger \ca$ are the lead and substrate electron number, respectively, with $\ca = u\hat{\gamma} + v \hat{\gamma}^\dagger$. With this transformation, the single-particle energies in the lead and substrate are measured from the chemical potential of the lead ($\mu_\mathrm{L}$) and substrate ($\mu_\mathrm{S}$), respectively, where the transformed Hamiltonian is
\begin{align}
\hat{\grave{H}} &= \hat{U} \hat{H}\hat{U}^\dagger - i \hbar \hat{U}\partial_\tau \hat{U}^\dagger\nonumber\\
&= (\HL - \mu_\mathrm{L} \hat{N}_\mathrm{L}) + (\HA - \mu_\mathrm{S} \hat{N}_\mathrm{S}) + \hat{U}\HT \hat{U}^\dagger,
\end{align}
with the tunneling Hamiltonian transformed as
\begin{align}
\hat{\grave{H}}_\mathrm{T} &= \hat{U} \HT\hat{U}^\dagger\nonumber\\
&= t e^{ieV\tau/\hbar} \hatcl^\dagger \ca + \mathrm{H.c.},
\end{align}
where $eV = \mu_{\mathrm{L}} - \mu_{\mathrm{S}}$.

The current operator is given by
\begin{equation}
\hat{I} = e\dotNL = ie[\NL,\hat{\grave{H}}_\mathrm{T}] = i\frac{e}{\hbar} \left(te^{ieV\tau/\hbar}\hatcl^\dagger \ca - \mathrm{H.c.} \right).
\end{equation}
By taking the expectation value of the current operator, we have
\begin{align}\label{eq:Itau}
I(\tau) &= \frac{e}{2h} \mathrm{Tr}\left\{\sigma_z[\gravet(\tau) \Glessabslead(\tau,\tau) - \Glessleadabs(\tau,\tau)\gravet^\dagger(\tau)] \right\},
\end{align}
where $\sigma_z$ is the $z$-Pauli Matrix in the Nambu basis.
In Eq.~\eqref{eq:Itau}, we have introduced the hopping matrix 
\begin{equation}
\gravet(\tau)  = e^{ieV\tau/\hbar\sigma_z}\check{t}= t \left(\begin{matrix} ue^{ieV\tau/\hbar} & ve^{ieV\tau/\hbar} \\ -v^*e^{-ieV\tau/\hbar} & -u^*e^{-ieV\tau/\hbar} \end{matrix} \right),
\end{equation}
and the lesser Green's function in the Nambu space [$(G^<_{\alpha \beta})_{ij} = i\langle \Psi^\dagger_{\beta j}\Psi_{\alpha i}\rangle$] with $i,j = \mathrm{L},\mathrm{A}$ denoting the quantities for the lead and ABS, respectively,  where $\Psi_{\mathrm{L}} = (\cl,\cl^\dagger)^T$ and $\Psi_{\mathrm{A}} = (\hatgamma,\hatgamma^\dagger)^T$. We can Fourier-expand the current and Green's functions in terms of the frequency $\omega_0 = eV/\hbar$, where we have
\begin{subequations}
\begin{align}
I(\tau) &= \sum_{n}I_ne^{in\omega_0\tau},\label{eq:currentFourier0} \\
G(\tau_1,\tau_2) &= \sum_n e^{in\omega_0\tau_2}\int_{-\infty}^{\infty}\frac{d\omega}{2\pi}e^{-i\omega(\tau_1-\tau_2)}G(\omega,\omega+n\omega_0).\label{eq:Gtau1tau2}
\end{align}
\end{subequations}
Let us denote $G_{mn}(\omega) \equiv G(\omega+m\omega_0,\omega+n\omega_0)$ for which we have $G_{mn}(\omega) = G_{m-n,0}(\omega+n\omega_0)$. 

The dc current which is the zeroth order ($I_0$) in the Fourier expansion of the current [Eq.~\eqref{eq:currentFourier0}] is given by
\begin{align}\label{eq:Izerocomp}
I_0 = \frac{e}{2h}t \int d\omega &\left(u \expomegap \GlessALeepo + v \expomegap \GlessALhepo + v^*\expomegam \GlessALehmo + u^*\expomegam \GlessALhhmo \right.\nonumber\\
 &\hspace{0.5 cm}\left.- u^*\expomegam\GlessLAeeop - v^*\expomegam\GlessLAehop -v\expomegap\GlessLAheom - u\expomegap \GlessLAhhom\right), 
\end{align}
where the superscripts $ee$, $eh$, $he$, and $hh$ denote the matrix elements in the Nambu space. Using the Langreth rule~\cite{haug2008quantum}
\begin{subequations}
\begin{align}
\Glessabslead &= \Grabs \gravet^\dagger\glesslead  + \Glessabs \gravet^\dagger\galead, \\
\Glessleadabs &= \grlead \gravet \Glessabs + \glesslead \gravet\Gaabs,
\end{align}
\end{subequations}
where $g_{\mathrm{L}} = \mathrm{diag} \left(g_{\mathrm{L}}^{ee},g_{\mathrm{L}}^{hh}\right)$, we have 
\begin{subequations}\label{eq:Glessmatrix}
\begin{align}
\GlessALeepo &= t\left[\left(\GrAeepp u^*  + \GrAehpp v^*\right) \expomegam \glessLeeoo + \left(\GlessAeepp u^*  + \GlessAehpp v^*\right) \expomegam \gaLeeoo  \right],\\
\GlessALhhmo &= -t\left[\left(\GrAhemm v  + \GrAhhmm u\right) \expomegap \glessLhhoo + \left(\GlessAhemm v  + \GlessAhhmm u\right) \expomegap \gaLhhoo  \right],\\
\GlessALhepo &= t\left[\left(\GrAhepp u^*  + \GrAhhpp v^*\right) \expomegam \glessLeeoo + \left(\GlessAhepp u^*  + \GlessAhhpp v^*\right) \expomegam \gaLeeoo  \right],\\
\GlessALehmo &= -t\left[\left(\GrAeemm v  + \GrAehmm u\right) \expomegap \glessLhhoo + \left(\GlessAeemm v  + \GlessAehmm u\right) \expomegap \gaLhhoo  \right],\\
\GlessLAeeop &= t\left[\glessLeeoo \expomegap \left(u\GaAeepp   + v\GaAhepp \right) + \grLeeoo \expomegap \left(u\GlessAeepp   + v\GlessAhepp \right)   \right],\\
\GlessLAhhom &= -t\left[ \glessLhhoo \expomegam \left(v^*\GaAehmm   + u^*\GaAhhmm \right)  + \grLhhoo \expomegam  \left(v^*\GlessAehmm   + u^*\GlessAhhmm \right)  \right],\\
\GlessLAheom &= -t\left[ \glessLhhoo \expomegam\left(v^*\GaAeemm   + u^*\GaAhemm \right)  + \grLhhoo\expomegam \left(v^*\GlessAeemm   + u^*\GlessAhemm \right)   \right],\\
\GlessLAehop &= t\left[\glessLeeoo\expomegap\left(u\GaAehpp   + v \GaAhhpp \right)   + \grLeeoo\expomegap \left(u\GlessAehpp   + v\GlessAhhpp \right)   \right].
\end{align}
\end{subequations}
Substituting Eq.~\eqref{eq:Glessmatrix} into Eq.~\eqref{eq:Izerocomp} and using $g^{ee}_\mathrm{L,00} = g^{hh}_\mathrm{L,00} = g_\mathrm{L}$, we have
\begin{align}
I_0 = \frac{e}{2h}t^2\int d\omega &\left\{ \left[|u|^2 (\Grabsee(\omega_+) - \Gaabsee(\omega_+)) + u v^* (\Grabseh(\omega_+) - \Gaabseh(\omega_+)) \right.\right.\nonumber\\
&\hspace{2 cm}\left.+ u^*v(\Grabshe(\omega_+)-\Gaabshe(\omega_+)) + |v|^2 (\Grabshh(\omega_+)-\Gaabshh(\omega_+))\right] \glesslead(\omega)\nonumber\\
&\hspace{0.2cm}- \left[|v|^2 (\Grabsee(\omega_-) - \Gaabsee(\omega_-)) + u v^* (\Grabseh(\omega_-) - \Gaabseh(\omega_-)) \right.\nonumber\\
&\hspace{2cm}\left.+ u^*v(\Grabshe(\omega_-)-\Gaabshe(\omega_-)) + |u|^2 (\Grabshh(\omega_-)-\Gaabshh(\omega_-)) \right] \glesslead(\omega)\nonumber\\
&\hspace{0.2cm} + \left[|u|^2 \Glessabsee(\omega_+) + u v^* \Glessabseh(\omega_+) + u^*v\Glessabshe(\omega_+) + |v|^2 \Glessabshh(\omega_+) \right] \left[\galead(\omega) - \grlead(\omega) \right] \nonumber\\
&\hspace{0.2cm} - \left.\left[|v|^2 \Glessabsee(\omega_-) + u v^* \Glessabseh(\omega_-) + u^*v\Glessabshe(\omega_-) + |u|^2 \Glessabshh(\omega_-) \right] \left[\galead(\omega) - \grlead(\omega) \right]\right\}.
\end{align}
Furthermore, by using the relation $G^< - G^> = G^a - G^r$, we obtain 
\begin{align}
I_0 = \frac{e}{2h}t^2\int d\omega &\left\{ \left[|u|^2 \Ggreatabsee(\omega)  + u v^* \Ggreatabseh(\omega) +u^*v \Ggreatabshe(\omega) + |v|^2 \Ggreatabshh(\omega)\right] \glesslead(\omega_-) \right. \nonumber\\
&\hspace{0.2cm} -\left[|v|^2 \Ggreatabsee(\omega)  + uv^* \Ggreatabseh(\omega) +u^*v \Ggreatabshe(\omega) + |u|^2 \Ggreatabshh(\omega)\right] \glesslead(\omega_+)\nonumber\\
&\hspace{0.2cm}-\left[|u|^2 \Glessabsee(\omega)  + u v^* \Glessabseh(\omega) +u^*v \Glessabshe(\omega) + |v|^2 \Glessabshh(\omega)\right] \ggreatlead(\omega_-) \nonumber\\
&\hspace{0.2cm} \left.+\left[|v|^2 \Glessabsee(\omega)  + u v^* \Glessabseh(\omega) +u^*v \Glessabshe(\omega) + |u|^2 \Glessabshh(\omega)\right] \ggreatlead(\omega_+)\right\}.
\end{align}
The current can be written more compactly as
\begin{align}\label{eq:dccurrenttot}
I &= \frac{e}{2h} \int d\omega\mathrm{Tr}\left[\Ggreatabs(\omega)\widetilde{\Sigma}_{\mathrm{A}}^{<}(\omega) - \Glessabs(\omega)\widetilde{\Sigma}_{\mathrm{A}}^{>}(\omega)  \right],
\end{align}
where
\begin{align}\label{eq:sigmalessgreat}
\widetilde{\Sigma}^{<,>}_{\mathrm{A}}(\omega) &= \checkt^\dagger \left(\begin{matrix}\glessgreatlead(\omega_-)& 0 \\0&-\glessgreatlead(\omega_+)\end{matrix}\right) \checkt \nonumber\\
&= t^2\left(\begin{matrix}u^* & -v \\ v^* & -u \end{matrix}\right)  \left(\begin{matrix}\glessgreatlead(\omega_-)& 0 \\0&-\glessgreatlead(\omega_+)\end{matrix}\right) \left(\begin{matrix}u & v \\ -v^* & -u^* \end{matrix}\right)\nonumber\\
&= t^2\left( \begin{matrix}|u|^2 \glessgreatlead(\omega_-)  - |v|^2 \glessgreatlead(\omega_+) & v u^*\left[ \glessgreatlead(\omega_-)  -  \glessgreatlead(\omega_+) \right] \\  u v^*\left[ \glessgreatlead(\omega_-)  -  \glessgreatlead(\omega_+) \right] & |v|^2 \glessgreatlead(\omega_-) - |u|^2 \glessgreatlead(\omega_+)
 \end{matrix} \right),
\end{align}
with $\glesslead (\omega)= f(\omega) (\galead(\omega)-\grlead(\omega))$, $\ggreatlead (\omega)= -(1-f(\omega)) (\galead(\omega)-\grlead(\omega))$ and  $g^r_{\mathrm{L}}(\omega_-) = g^r_{\mathrm{L}}(\omega_+)\equiv -i\pi\nu_0$, where where $\nu_0$ is the density of states at the lead Fermi energy.
Substituting the expressions for $\widetilde{\Sigma}^{<,>}_{\mathrm{A}}(\omega)$ [Eqs.~\eqref{eq:sigmalessgreat}],  $\Glessabs(\omega)$ [Eqs.~\eqref{eq:Glessall}], and the corresponding equations for $\Ggreatabs(\omega)$ into Eq.~\eqref{eq:dccurrenttot}, we obtain Eq.~\eqref{eq:IV} of the main text:
\begin{equation}\label{eq:IVapp}
 I(V) = \frac{e}{h}\Gamma^2\int d\omega \mathcal{A}(\omega)[f(\omega_-)-f(\omega_+)],
\end{equation}
where
\begin{equation}
\mathcal{A}(\omega) = \frac{4 |uv|^2}{\left[\omega  -\frac{(\eabs+\sqrt{2}\lambda \avex)^2}{\omega}- \frac{(\Gamma_u - \Gamma_v)^2}{4\omega}\right]^2+ (\Gamma_u + \Gamma_v)^2},
\end{equation}
with $\Gamma_u  = \Gamma |u|^2$ and $\Gamma_v  =  \Gamma |v|^2$. Note that the term $\avex$ is evaluated self-consistently using
\begin{align}\label{eq:selfconstsuppl}
\avex &= -\frac{\sqrt{2}\lambda}{ \Omega} \langle \hatgamma^\dagger(\tau) \hatgamma (\tau) \rangle = i\frac{\sqrt{2}\lambda}{ \Omega}\int \frac{d\omega}{2\pi}\Glessabsee(\omega),
\end{align}
or Eq.~\eqref{eq:selfconst} of the main text:
\begin{align}\label{eq:selfconstsuppl}
\avex &= -\frac{\sqrt{2}\lambda}{ \Omega} \langle \hatgamma^\dagger (\tau) \hatgamma (\tau) \rangle = -\frac{\lambda}{\sqrt{2} \Omega}\left(1+ \langle \hatgamma^\dagger (\tau) \hatgamma (\tau) \rangle - \langle  \hatgamma (\tau) \hatgamma^\dagger (\tau)\rangle\right) = -\frac{\lambda}{\sqrt{2}\Omega}\left\{1-i\int \frac{d\omega}{2\pi} \mathrm{Tr}\left[\Glessabs(\omega)\sigma_z\right]\right\}.
\end{align}

\section{Explicit expressions for $\Glessgreatabs(\omega)$}\label{sec:tbh}
In this section, we evaluate the expressions for the ABS lesser and greater Green's functions $\Glessgreatabs(\omega)$ which are used to calculate the current [Eq.~\eqref{eq:IV} of the main text] and the expectation value of the boson displacement operator $\langle \hat{x} \rangle$ [Eq.~\eqref{eq:selfconst} of the main text]. Using the Fourier expansion as in Eq.~\eqref{eq:Gtau1tau2}, we can relate the ABS lesser and greater Green's function in the frequency domain $\Glessgreatabs(\omega)$ to their time-domain counterparts~\cite{haug2008quantum}, i.e.,
\begin{subequations}
\begin{align}
 \Glessabs(\tau_1,\tau_2) &= \left(\begin{matrix} \Glessabsee(\tau_1,\tau_2) & \Glessabseh(\tau_1,\tau_2) \\ \Glessabshe(\tau_1,\tau_2) & \Glessabshh(\tau_1,\tau_2)\end{matrix}\right) \equiv i \left(\begin{matrix} \langle \gamma^\dagger(\tau_2) \gamma (\tau_1)\rangle & \langle \gamma (\tau_2)\gamma(\tau_1) \rangle \\ \langle\gamma^\dagger(\tau_2) \gamma^\dagger (\tau_1)\rangle & \langle\gamma (\tau_2)\gamma^\dagger (\tau_1) \rangle \end{matrix}\right),\\
 \Ggreatabs(\tau_1,\tau_2) &= \left(\begin{matrix} \Ggreatabsee(\tau_1,\tau_2) & \Ggreatabseh(\tau_1,\tau_2) \\ \Ggreatabshe(\tau_1,\tau_2) & \Ggreatabshh(\tau_1,\tau_2)\end{matrix}\right) \equiv -i \left(\begin{matrix} \langle \gamma(\tau_1) \gamma^\dagger (\tau_2)\rangle & \langle \gamma^\dagger (\tau_1)\gamma^\dagger(\tau_2) \rangle \\ \langle\gamma(\tau_1) \gamma (\tau_2)\rangle & \langle\gamma^\dagger (\tau_1)\gamma (\tau_2) \rangle \end{matrix}\right).
\end{align}
\end{subequations}

To evaluate $\Glessgreatabs(\omega)$, we begin by writing the ABS Green's function in the Lehmann representation as
\begin{align}\label{eq:gomega}
g_\mathrm{A}(\omega) &= \frac{\Phi_+ \Phi^\dagger_+}{\omega - (\eabs+\sqrt{2}\lambda \avex)} +  \frac{\Phi_- \Phi^\dagger_-}{\omega + (\eabs+\sqrt{2}\lambda\avex)}.
\end{align}
where $\Phi_+ = (1, 0)^T$ and $\Phi_- = (0, 1)^T$ are the positive- and negative-energy eigenfunction of the ABS written in the Nambu basis $(\hatgamma,\hatgamma^\dagger)^T$.
The ABS's self energy due to the lead coupling is $\Sigmarabs(\omega)  =\checkt^\dagger \mathrm{diag} (g_L^r(\omega_-),g_L^r(\omega_+)) \checkt$ where $\checkt = t \left(\begin{matrix} u & v \\ -v^* & -u^* \end{matrix} \right)$ is the hopping matrix, and $g^r_{\mathrm{L}}(\omega_-) = g^r_{\mathrm{L}}(\omega_+)= -i\pi\nu_0$ is the lead retarded Green's function with $\omega_\pm = \omega \pm eV$. Similar relations apply for $\Sigma_{\mathrm{A}}^{a,<,>}(\omega)$. The ABS lesser Green’s
function is~\cite{haug2008quantum}
\begin{align}~\label{eq:glessabs}
\Glessabs(\omega) &= \glessabs(\omega) + \grabs(\omega) \Sigmarabs(\omega) \Glessabs(\omega) +\left[\grabs(\omega) \Sigmalessabs(\omega) +\glessabs(\omega) \Sigmaaabs(\omega)\right]\Gaabs(\omega) \nonumber\\
&= \frac{1}{1- \grabs(\omega) \Sigmarabs(\omega) }\left[ \glessabs(\omega)\left(1+\Sigmaaabs(\omega) \Gaabs(\omega) \right) + \grabs(\omega) \Sigmalessabs(\omega) \Gaabs(\omega)\right],
\end{align}
where $\Graabs = \graabs(1-\graabs\Sigmaraabs)^{-1}$, $g^<_{j} (\omega)= f(\omega) (g^a_j-g^r_j)$ and $g^>_j (\omega)= -(1-f(\omega)) (g^a_j-g^r_j)$ with $j$ = L, A. The explicit expressions of the matrix elements of $\Glessabs(\omega) = \left(\begin{matrix} \Glessabsee(\omega) & \Glessabseh(\omega) \\ \Glessabshe(\omega) & \Glessabshh(\omega)\end{matrix}\right)$ can be evaluated as 
\begin{subequations}\label{eq:Glessall}
\begin{align}\label{eq:Gless}
\Glessabsee(\omega) &= \frac{i}{D} \left\{ \left[\Gamma_uf(\omega_-) +\Gamma_vf(\omega_+)\right] \left[ \left(\omega + \eabs+\sqrt{2}\lambda \avex\right)^2 + \left(\frac{\Gamma_u -\Gamma_v}{2}  \right)^2  \right] \right\},\\
\Glessabseh(\omega) &= \frac{i}{D} \left\{\Gamma u^* v \left[\left(f(\omega_+)+f(\omega_-)\right)\left(\omega^2 - (\eabs+\sqrt{2}\lambda \avex)^2 - \left(\frac{\Gamma_u -\Gamma_v}{2}  \right)^2\right)  + i\omega\left(f(\omega_+)-f(\omega_-)\right)\left(\Gamma_u -\Gamma_v \right) \right]\right\},\\
\Glessabshe(\omega) &= \frac{i}{D} \left\{\Gamma u v^*\left[\left(f(\omega_+)+f(\omega_-)\right) \left(\omega^2 - (\eabs+\sqrt{2}\lambda \avex)^2 - \left(\frac{\Gamma_u -\Gamma_v}{2}  \right)^2\right) - i\omega\left(f(\omega_+)-f(\omega_-)\right)\left(\Gamma_u -\Gamma_v \right) \right] \right\},\\
\Glessabshh(\omega) &=  \frac{i}{D} \left\{ \left[\Gamma_uf(\omega_+) +\Gamma_vf(\omega_-)\right] \left[ \left(\omega - (\eabs+\sqrt{2}\lambda \avex)\right)^2 + \left(\frac{\Gamma_u -\Gamma_v}{2}  \right)^2  \right] \right\},
\end{align}
\end{subequations}
where $\omega_{\pm} = \omega \pm eV$ and
\begin{align}
D &=\left[\omega^2 - (\eabs+\sqrt{2}\lambda \avex)^2 - \frac{(\Gamma_u - \Gamma_v)^2}{4} \right]^2+ \omega^2(\Gamma_u+\Gamma_v)^2,
\end{align}
with $\Gamma_u = \Gamma|u|^2$ and $\Gamma_v = \Gamma|v|^2$. The expressions for the matrix elements of the ABS greater Green's function $\Ggreatabs(\omega)$ can be obtained from Eq.~\eqref{eq:Glessall} by using the substitutions: $f(\omega_-) \rightarrow f(\omega_-) - 1$ and $f(\omega_+) \rightarrow f(\omega_+) - 1$.

\section{Dependence of the current and conductance calculated from Keldysh approach on ABS-boson coupling strength, temperature, boson frequency, and lead-tunnel coupling strength}

\begin{figure}[h!]
\centering
\includegraphics[width=\linewidth]{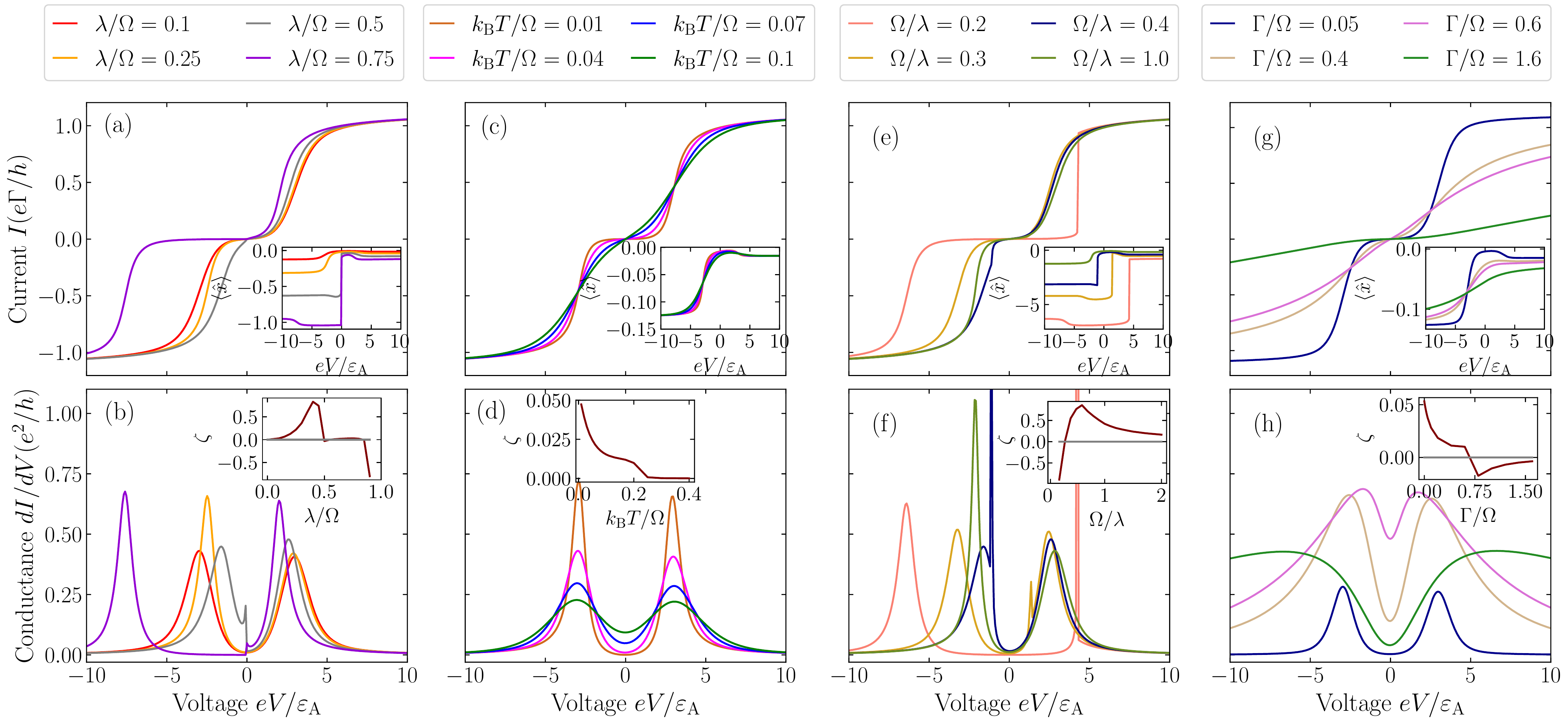}
\caption{Current $I$ (Upper panels) and conductance $dI/dV$ (Lower panels) of boson-coupled ABSs vs voltage $V$ calculated using the Keldysh approach [Eq.~\eqref{eq:IV} of the main text] subject to the self-consistency condition [Eq.~\eqref{eq:selfconst} of the main text]. The plots are for (a,b) different ABS-boson coupling strengths $\lambda$ with $\kb T = 0.4$, $\eabs = 3.0$, and $\Gamma = 1.0$, (c,d) different temperatures $T$ with $\lambda = 1.0$, $\eabs = 3.0$, and $\Gamma=  1.0$, (e,f) different boson frequencies $\Omega$ with  $\kb T = 0.4$, $\eabs = 3.0$, $\lambda = 1.0$ and $\Gamma =1.0$, and (g,h) different lead-tunnel coupling strengths $\Gamma$ with $\kb T = 0.4$, $\lambda=1.0$ and $\eabs = 3.0$. Inset in upper panels: (a) Mean-field boson displacement value $\avex$ vs $\lambda/\Omega$, (c) $\avex$ vs temperature $T$, (e) $\avex$ vs boson frequency $\Omega$, and (g) $\avex$ vs  $\Gamma/\Omega$. Inset in lower panels: (b) Conductance PH asymmetry $\zeta$ vs $\lambda/\Omega$,  (d) $\zeta$ vs temperature $T$,  (f) $\zeta$ vs boson frequency $\Omega$, and (h) $\zeta$ vs $\Gamma/\Omega$. The parameters used for all panels are: $|u|^2/|v|^2 = 1/9$. Note that for panels (a,b,c,d,g,h), we use high-frequency bosons ($\Omega = 10$) where $\Omega > 2 \eabs+k_{\mathrm{B}}T$.}\label{fig:currentKeldyshappendix}
\end{figure}

Figure~\ref{fig:currentKeldyshappendix} shows the current (upper panels) and conductance (lower panels) of boson-coupled ABSs calculated from the Keldysh approach~[Eq.~\eqref{eq:IV} of the main text] subject to the self-consistency condition [Eq.~\eqref{eq:selfconst} of the main text]. The plots are shown for different ABS-boson coupling strengths $\lambda$ [Figs.~\ref{fig:currentKeldyshappendix}(a,b)], temperatures $T$ [Figs.~\ref{fig:currentKeldyshappendix}(c,d)], boson frequencies $\Omega$ [Figs.~\ref{fig:currentKeldyshappendix}(e,f)] and lead-tunnel coupling strengths $\Gamma$ [Figs.~\ref{fig:currentKeldyshappendix}(g,h)]. While in Fig.~\ref{fig:Keldyshcurrent} of the main text, we have shown that the PHS breaking holds for the case of low-frequency bosons, here we will show that it also holds for the case of high-frequency bosons, i.e., $\Omega > 2 \eabs + k_{\mathrm{B}}T$. Unlike the perturbative calculation in the rate equation, the PHS breaking calculated from the Keldysh approach arises due to non-perturbative effects of tunneling, i.e., the PH asymmetry of the mean-field boson displacement value $\avex$. In the non-perturbative regime,  electrons can tunnel from the lead into virtual states in the superconductor by emitting or absorbing bosons with high frequencies where energy violation is allowed for sufficiently large tunnel coupling ($\Gamma \gtrsim \Omega$), resulting in PHS breaking of subgap conductances. This energy violation is allowed as long as the energy violation in the first tunneling process is negated by the second tunneling process which the conserves the total energy of a full cycle of transferring a pair of electrons in the two-step tunneling process.

Figure~\ref{fig:currentKeldyshappendix}(b) shows that the magnitude of the conductance PH asymmetry $\zeta$ has a nonmonotonic dependence on the ABS-boson coupling strength $\lambda$. The conductance PH asymmetry $\zeta$ first increases with increasing $\lambda$ where the two peaks approach each other until they reach a certain minimum distance. Note that for this range of  $\lambda$, the higher peak is at positive voltage for the case where $|u|^2 > |v|^2$ while for the case where $|v|^2 > |u|^2$, the higher peak is at negative voltage.  After the PH asymmetry reaches a maximum, it decreases to zero and stays there for a range of $\lambda$ where the peaks remain more or less at the same place. As $\lambda$ increases and reaches a certain value, the high and low peaks switch positions, i.e., from negative to positive voltage and vice versa. As $\lambda$ keeps increasing, the two peaks move away from each other and the magnitude of the PH asymmetry increases. For large enough $\lambda$, the positions of the conductance peaks are no longer PH symmetric [see purple curve in Fig.~\ref{fig:currentKeldyshappendix}(b)]. Note that the results for large $\lambda$ may not be reliable as our mean-field treatment of interactions may break down in this regime. 

Figure~\ref{fig:currentKeldyshappendix}(d) shows that the conductance PH asymmetry $\zeta$ decreases with increasing temperature $T$ due the temperature broadening of the conductance peaks.  The dependence of the ABS conductance on the boson frequency $\Omega$ is shown in Fig.~\ref{fig:currentKeldyshappendix}(f). The PH asymmetry $\zeta$ has a nonmonotonic behavior with the boson frequency $\Omega$ where its magnitude first decreases to zero with increasing $\Omega$. This corresponds to the two conductance peaks moving towards each other as $\Omega$ increases which is due to the decrease in the effective ABS-boson coupling strength $\lambda/\Omega$. After the PH asymmetry reaches zero, it switches sign which corresponds to the high and low peaks switching sides. As $\Omega$ increases, the two peaks move towards each other and the PH asymmetry increases to a certain maximum value. Having reached its maximum, the PH asymmetry $\zeta$ decreases with increasing $\Omega$ which corresponds to the decrease in  the effective ABS-boson coupling strength $\lambda/\Omega$. Figure~\ref{fig:currentKeldyshappendix}(h) shows the dependence of the conductance on the lead tunnel coupling $\Gamma$. As shown in the inset of panel (h), the PH asymmetry $\zeta$ has a non-monotonic dependence on the lead-tunnel coupling $\Gamma$ where it first decreases as $\Gamma$ increases. After the PH asymmetry reaches zero, it changes sign and increases in magnitude to a certain maximum value as $\Gamma$ increases. Having reached its maximum, the PH asymmetry then decreases as $\Gamma$ increases. Note that unlike the rate equation, our mean-field Keldysh approach shows that the conductance in the tunneling limit ($\Gamma/\Omega \ll 1$) still exhibits PH asymmetry even for high-frequency bosons. Since the treatment of interactions within the rate equation is exact in the tunneling limit, our Keldysh results obtained using the mean-field treatment of interactions may not be correct in this tunneling limit. This is because the mean-field approximation breaks down in this limit due to the singularity in the tunneling density of states. For the case where the tunnel coupling is not too small, the mean-field approximation is valid and we can see from Fig.~\ref{fig:currentKeldyshappendix}(d) that unlike the rate equation, that subgap conductance calculated from the Keldysh approach can still be PH asymmetric for high-frequency boson case.  Finally we note that since we ignore the Fock term in the mean-field approximation, the conductance calculated from the Keldysh approach has no boson sidebands.

\section{Details on Model II. boson-assisted tunneling model into ABS}\label{sec:indirectHABS}
In this section, we consider a boson-assisted tunneling Hamiltonian of the form
\begin{equation}\label{eq:Htunnelindirect}
\HT= t(\hatb + \hatb^\dagger)\hatcl^\dagger\ca + \mathrm{H.c.}
\end{equation}
This tunneling Hamiltonian can be obtained by first projecting the microscopic  Hamiltonian [Eq.~\eqref{eq:Hephmicro}] onto the lowest and second-lowest energy sector $\alpha, \beta = 1,2$ and followed by integrating out the second-lowest Bogoliubov operator $\gamma_2$ from the total Hamiltonian of the system. 

Projecting the ABS and tunneling Hamiltonian onto the lowest and second-lowest energy state gives
\begin{subequations}
\begin{align}
\hat{H}_{\mathrm{A}} &= \sum_{\alpha=1}^2  \varepsilon_\alpha \gamma^\dagger_{\alpha}\gamma_\alpha + \sum_{\alpha,\beta=1}^2 \biggl(\lambdatc_{\alpha\beta}\gamma^\dagger_{\alpha}\gamma_{\beta}+\lambdatd_{\alpha\beta}\gamma_{\alpha}\gamma_{\beta}+\mathrm{H.c.}\biggr)( \hat{b}^\dagger+\hat{b}) + \Omega (\hatb^\dagger + \hatb),\label{eq:Hephtwo}\\
\hat{H}_{\mathrm{T}} &= \sum_{\alpha=1}^2 t_\alpha \cl^\dagger \biggl(\sum_{\beta=1}^{2} u_{\alpha \beta}(0)\hatgamma_\beta + v_{\alpha\beta}(0)\hatgamma^\dagger_\beta\biggr) + \mathrm{H.c.}
\end{align}
\end{subequations}
For simplicity, we will choose parameters such that the tunneling term into the lowest Bogoliubov (ABS) operator ($\gamma_1$ and $\gamma_1^\dagger$) vanishes where we only have the tunneling into the second lowest Bogoliubov operator ($\gamma_2$ and $\gamma_2^\dagger$), i.e.,
\begin{align}
\hat{H}_{\mathrm{T}}&=  \ttilde \cl^\dagger \biggl(\utilde\hatgamma_2 + \vtilde\hatgamma^\dagger_2\biggr) + \mathrm{H.c.}\label{eq:HTgammatwo}
\end{align}
Note that in the above, we have defined $\tilde{t} \tilde{u}\equiv t_2u_{22}(0)+ t_1u_{12}(0)$, $\tilde{t} \tilde{v}\equiv t_2v_{22}(0)+ t_1v_{12}(0)$, and we have also chosen parameters such that the tunneling term into the ABS vanishes, i.e., $t_2v_{21}(0)+ t_1v_{11}(0) = 0$ and $t_2u_{21}(0)+ t_1u_{11}(0) = 0$. We note that we choose these parameters only for simplicity and our results on PHS breaking hold in general even without this simplification.

In the following, we will see that the tunneling Hamiltonian [Eq.~\eqref{eq:HTgammatwo}] and the electron-boson interaction Hamiltonian [Eq.~\eqref{eq:Hephtwo}] give rise to a boson-assisted tunneling into the lowest energy ABS. To this end, we will use the path integral formalism to integrate out $\gamma_2$. We begin by writing down the partition function of the system as
\begin{align}\label{eq:partition}
Z &= \int_{-\infty}^{\infty} \prod_{\alpha=1}^2 \mathcal{D}\gbar_\alpha\mathcal{D} \gamma_\alpha \mathcal{D} \clkbar \mathcal{D}\clk    \mathcal{D}\bar{b}\mathcal{D}b\biggl(\mathrm{exp}[i(S_{\alpha}+S_{\mathrm{b}}+S_\mathrm{L}+S_{\text{e-b}}+S_\mathrm{T})]\biggr)\nonumber\\
&= \int_{-\infty}^{\infty} \prod_{\alpha=1}^2 \mathcal{D}\gbar_\alpha\mathcal{D} \gamma_\alpha \mathcal{D} \clkbar\mathcal{D}\clk \mathcal{D}\bar{b}\mathcal{D}b  \biggl(\mathrm{exp}\bigg[i\int_{-\infty}^{\infty} d\tau \bigg(\hbar \gbar_\alpha\partial_\tau \gamma_\alpha + \hbar \bar{b}\partial_\tau b +\hbar \clkbar\partial_\tau \clk  - \varepsilon_{\alpha} \gbar_\alpha \gamma_\alpha - \Omega \bar{b}b \nonumber\\
&\qquad\qquad\qquad\qquad - \sum_k\elk \clkbar \clk-\sum_{\beta=1,2}(\lambdatc_{\alpha\beta}\bar{\gamma}_{\alpha}\gamma_{\beta}+\lambdatd_{\alpha\beta}\gamma_{\alpha}\gamma_{\beta}+\mathrm{h.c.})( \bar{b}+b) - \ttilde( \utilde \clbar\gamma_2 + \vtilde \clbar \gbar_2 + \mathrm{H.c.})\biggr)\biggr] \biggr),
\end{align}
where the actions are given by
\begin{subequations}
\begin{align}
S_\alpha &= \int d\tau  \bar{\gamma}_\alpha(\tau) (\hbar\partial_\tau - \varepsilon_\alpha)\gamma_\alpha(\tau),\\
S_{\mathrm{b}} &= \int d\tau \bar{b}(\tau) (\hbar\partial_\tau - \Omega)b(\tau),\\
S_{\text{e-b}} &= -\int d\tau \sum_{\alpha,\beta=1}^2(\lambdatc_{\alpha\beta}\bar{\gamma}_{\alpha}(\tau)\gamma_{\beta}(\tau)+\lambdatd_{\alpha\beta}\gamma_{\alpha}(\tau)\gamma_{\beta}(\tau)+\mathrm{H.c.})( \bar{b}(\tau)+b(\tau)),\\
S_\mathrm{L} &= \sum_k\int d\tau  \bar{c}_{\mathrm{L},k}(\tau) (\hbar\partial_\tau -\elk)\clk(\tau),\\
S_{\mathrm{T}} &= -\int d\tau \biggl[\ttilde( \utilde\clbar(\tau)\gamma_2(\tau) + \vtilde\clbar(\tau)\gbar_2(\tau)) + \mathrm{H.c.}\biggr].
\end{align}
\end{subequations}
We assume $\lambda_{22}\ll \varepsilon_2$, so that we can integrate out $\gamma_2$ and $\gbar_2$ by using the Gaussian integral 
\begin{align}\label{eq:gaussian}
&\int_{-\infty}^{\infty}\mathcal{D}\gamma_2 \mathcal{D}\gbar_2 \exp \biggl\{i\varepsilon_2\gbar_2 \gamma_2 -i \biggl[ \bigl(\lambdatc_{12}\gbar_1+\lambdatd_{12}\gamma_1   + \ttilde (\utilde \clbar + \vtilde^* \cl )\bigr)\gamma_2(b + \bbar) + \mathrm{H.c.}\biggr]\biggr\}\nonumber\\
&= \pi\exp\bigl\{-(b+ \bbar) \bigl( \lambdatc_{12}\gbar_1(\tau')+\lambdatd_{12}\gamma_1(\tau')  + \ttilde (\utilde \clbar(\tau') + \vtilde^* \cl(\tau') )\bigr) i(\Gttt(\tau-\tau'))^{-1}\nonumber\\
&\qquad\qquad\qquad\times(b+ \bbar)\bigl(  (\lambdatc_{12})^*\gamma_1+(\lambdatd_{12})^* \gbar_1(\tau) +  \ttilde (\utilde^* \cl(\tau) + \vtilde \clbar(\tau) )\bigr)\bigr\},
\end{align}
Substituting Eq.~\eqref{eq:gaussian} into Eq.~\eqref{eq:partition} and ignoring the $\lambda_{22}$ term, we have
\begin{align}
Z &= \mathrm{const}\times\int_{-\infty}^{\infty} \mathcal{D}\gbar_1 \mathcal{D}\gamma_1 \mathcal{D}\clkbar \mathcal{D}\clk\mathcal{D}\bar{b}\mathcal{D}b \nonumber\\
&\times\mathrm{exp}\biggl\{i\int_{-\infty}^{\infty} d\tau \bigg[\hbar \gbar_1\partial_\tau \gamma_1 +\hbar \bbar\partial_\tau b +\hbar \clkbar \partial_\tau \clk - \varepsilon_{1} \gbar_1 \gamma_1 - \Omega\bbar b - 2\lambdatc_{11} \gbar_1 \gamma_1 (\bbar+b)-\chi(\bbar+b)- \sum_k\elk \clkbar \clk \nonumber\\
&\qquad\qquad\qquad   -\int_{-\infty}^{\infty} d\tau' \bigg( (b+ \bbar) \bigl[ \lambdatc_{12}\gbar_1(\tau')+\lambdatd_{12}\gamma_1(\tau')  + \ttilde (\utilde \clbar(\tau') + \vtilde^* \cl(\tau') )\bigr]\bigg) i(\Gttt(\tau-\tau'))^{-1}\nonumber\\
&\qquad\qquad\qquad\qquad\qquad\times(b+ \bbar)\bigl[(\lambdatc_{12})^*\gamma_1(\tau) + (\lambdatd_{12})^* \gbar_1(\tau) +  \ttilde (\utilde^* \cl(\tau) + \vtilde \clbar(\tau) )\biggr]\biggr\},
\end{align}
where $\chi =\sum_{lm}\int dx dx' g_{lm}(x,x')$ and
$\Gaat$ is the time-ordered Green's function~\cite{Kamenev2010Keldysh} defined by
\begin{equation}
  i\Gaat(\tau - \tau') \equiv \langle\gamma_\alpha(\tau)\gamma_\alpha(\tau') \rangle  = \theta(\tau - \tau')i \Gaa^>(\tau-\tau') + \theta(\tau'-\tau)i\Gaa^<(\tau-\tau'),
\end{equation}
with
\begin{subequations}
\begin{align}
i\Gaa^<(\tau - \tau') &= -f(\varepsilon_\alpha)\exp(-i\varepsilon_{\alpha}(\tau-\tau')),\\
i\Gaa^>(\tau - \tau') &= (1-f(\varepsilon_\alpha))\exp(-i\varepsilon_{\alpha}(\tau-\tau')),
\end{align}
\end{subequations}
and $\theta(\tau)$ being the Heaviside step function. 
We consider $\varepsilon_2 \gg \kb T$, where we have the Fermi function $f(\varepsilon_2) = 1$ which gives $iG_{22}(\tau - \tau') = -\theta(\tau'-\tau)\exp(-i\varepsilon_{2}(\tau-\tau'))$. Using this, we can then write the partition function as
\begin{align}
Z &=\mathrm{const}\times\int_{-\infty}^{\infty} \mathcal{D}\gbar_1 \mathcal{D}\gamma_1 \mathcal{D}\clkbar \mathcal{D}\clk\mathcal{D}\bbar\mathcal{D}b \nonumber\\
&\times\mathrm{exp}\bigg\{i\int_{-\infty}^{\infty} d\tau \biggl[\hbar \gbar_1\partial_\tau \gamma_1 +\hbar \bbar\partial_\tau b+\hbar \clbar \partial_\tau \cl - \varepsilon_{1} \gbar_1 \gamma_1-\Omega \bbar b- 2\lambda_{11}^{(2)} \gbar_1 \gamma_1 (\bar{b}+b)-\chi(\bar{b}+b)- \sum_k\elk \clkbar \clk \biggr] \nonumber\\
&\qquad\qquad\qquad    +\int_{-\infty}^{\infty}d\tau\int_{\tau}^{\infty} d\tau' \biggl[\bigg( (b + \bar{b}) \bigl[\lambdatd_{12}\gamma_1(\tau')  + \lambdatc_{12}\gbar_1(\tau') + \ttilde (\utilde \clbar(\tau') + \vtilde^* \cl(\tau') )\bigr]\bigg) \exp(i\varepsilon_{2}(\tau-\tau'))\nonumber\\
&\qquad\qquad\qquad\qquad\qquad\times(b + \bbar)\bigl[(\lambdatc_{12})^*\gamma_1(\tau) + (\lambdatd_{12})^* \gbar_1(\tau) +  \ttilde (\utilde^* \cl(\tau) + \vtilde \clbar(\tau) )\biggr]\biggr\}.
\end{align}
Defining $\tau_1 \equiv (\tau+\tau')/2$ and $\tau_2 \equiv \tau'- \tau$, we have
$\int_{-\infty}^{\infty} d\tau \int_{\tau}^{\infty} d\tau'   = \int_{-\infty}^{\infty} d\tau_1 \int_{0}^{\infty} d\tau_2 $ and 
\begin{align}
Z &= \mathrm{const}\times\int_{-\infty}^{\infty} \mathcal{D}\gbar_1 \mathcal{D}\gamma_1 \mathcal{D}\clkbar \mathcal{D}\clk\mathcal{D}\bbar \mathcal{D}b \nonumber\\
&\times\mathrm{exp}\bigg\{i\int_{-\infty}^{\infty} d\tau \biggl[\hbar \gbar_1\partial_\tau \gamma_1 +\hbar \bbar \partial_\tau b +\hbar \clbar \partial_\tau \hatclk - \varepsilon_{1} \gbar_1 \gamma_1-\Omega \bbar b- 2\lambdatc_{11} \gbar_1 \gamma_1 (\bar{b}+b)-\chi(\bar{b}+b)- \sum_k\elk \clkbar \clk \biggr] \nonumber\\
&\qquad\qquad\qquad    +\int_{-\infty}^{\infty}d\tau_1\int_{0}^{\infty} d\tau_2 \biggl[\bigg( (b + \bar{b}) \bigl[\lambdatd_{12}\gamma_1(\tau_1+\frac{\tau_2}{2})  + \lambdatc_{12}\gbar_1(\tau_1+\frac{\tau_2}{2}) + \ttilde (\utilde \clbar(\tau_1+\frac{\tau_2}{2}) + \vtilde^* \cl(\tau_1+\frac{\tau_2}{2}) )\bigr]\bigg) \nonumber\\
&\qquad\qquad\qquad\times\exp(-i\varepsilon_{2}\tau_2)(b + \bbar)\bigl[(\lambdatc_{12})^*\gamma_1(\tau_1-\frac{\tau_2}{2}) + (\lambdatd_{12})^* \gbar_1(\tau_1-\frac{\tau_2}{2}) +  \ttilde (\utilde^* \cl(\tau_1-\frac{\tau_2}{2}) + \vtilde \clbar(\tau_1-\frac{\tau_2}{2}) )\bigr]\biggr]\biggr\}.
\end{align}
Assuming a slow variation of $\cl$, $\clbar$, $\gamma_1$ and $\bar{\gamma}_1$, we obtain
\begin{align}
&\int_{-\infty}^{\infty}d\tau_1\int_{0}^{\infty} d\tau_2 \biggl\{\biggl[\ttilde (\utilde \clbar(\tau_1+\frac{\tau_2}{2}) + \vtilde^* \cl(\tau_1+\frac{\tau_2}{2}) )\exp(-i\varepsilon_{2}\tau_2)(b + \bbar)\bigl[(\lambdatc_{12})^*\gamma_1(\tau_1-\frac{\tau_2}{2}) + (\lambdatd_{12})^* \gbar_1(\tau_1-\frac{\tau_2}{2}) \bigr]\biggr]\nonumber\\
&\qquad\qquad\qquad\qquad+\biggl[(b + \bbar) \bigl[ \lambdatc_{12}\gbar_1(\tau_1+\frac{\tau_2}{2}) + \lambdatd_{12}\gamma_1(\tau_1+\frac{\tau_2}{2})\bigr] \exp(-i\varepsilon_{2}\tau_2)\ttilde (\utilde^* \cl(\tau_1-\frac{\tau_2}{2}) + \vtilde \clbar(\tau_1-\frac{\tau_2}{2}) )\biggr]\biggr\}\nonumber\\
&\approx-i\int_{-\infty}^{\infty}d\tau_1 \biggl\{ \frac{\ttilde}{\varepsilon_2}\biggl[ (\utilde \clbar(\tau_1) + \vtilde^* \cl(\tau_1))(b + \bbar)\bigl[(\lambdatc_{12})^*\gamma_1(\tau_1) + (\lambdatd_{12})^* \gbar_1(\tau_1) \bigr]\biggr] + \mathrm{H.c.}\biggr\},
\end{align}
and
\begin{align}\label{eq:direct}
&\int_{-\infty}^{\infty}d\tau_1\int_{0}^{\infty} d\tau_2 \biggl\{(b + \bbar)^2\biggl[\lambdatd_{12}\gamma_1(\tau_1+\frac{\tau_2}{2})  + \lambdatc_{12}\gbar_1(\tau_1+\frac{\tau_2}{2})\exp(-i\varepsilon_{2}\tau_2)\bigl[(\lambdatc_{12})^*\gamma_1(\tau_1-\frac{\tau_2}{2}) + (\lambdatd_{12})^* \gbar_1(\tau_1-\frac{\tau_2}{2}) \bigr]\biggr]\nonumber\\
&\qquad\qquad\qquad\qquad+\biggl[\ttilde (\utilde \clbar(\tau_1+\frac{\tau_2}{2}) + \vtilde^* \cl(\tau_1+\frac{\tau_2}{2}) )\exp(-i\varepsilon_{2}\tau_2)\ttilde (\utilde^* \cl(\tau_1-\frac{\tau_2}{2}) + \vtilde \clbar(\tau_1-\frac{\tau_2}{2}) )\biggr]\biggr\}\nonumber\\
&=-i\int_{-\infty}^{\infty}d\tau_1 \biggl\{ \frac{(b + \bbar)^2}{\varepsilon_2}\biggl[ \biggl( |\lambdatc_{12}|^2-  |\lambdatd_{12}|^2\biggr) \gbar_1\gamma_1 + |\lambdatd_{12}|^2\biggr]+\frac{|\ttilde|^2}{\varepsilon_2}\biggl[ \biggl((|\utilde|^2- |\vtilde|^2 )\clbar(\tau_1)\cl(\tau_1) + |\vtilde|^2 \biggr)\biggr] \biggr\},
\end{align}
where we have ignored the boundary term at $\tau_2 = \infty$ since it is highly oscillating and thus averages to zero. The terms containing $(b + \bbar)^2$ in Eq.~\eqref{eq:direct} can be ignored since they are of second order in $\lambda_{12}^{(c,d)}$ where we assume $\lambda_{12}^{(c,d)}/\varepsilon_2 \ll 1$. The  terms proportional to $|\ttilde|^2$ renormalize the lead electrons' energies as well as their wave functions and can thus be subsumed into the lead Hamiltonian $\HL$. As a result, we have
\begin{align}\label{eq:Zeff}
Z &= \mathrm{const}\times\int_{-\infty}^{\infty} \mathcal{D}\gbar_1 \mathcal{D}\gamma_1 \mathcal{D}\clkbar \mathcal{D}\clk\mathcal{D}\bbar\mathcal{D}b \nonumber\\
&\times\mathrm{exp}\bigg\{i\int_{-\infty}^{\infty} d\tau \biggl[\hbar \gbar_1\partial_\tau \gamma_1 +\hbar \bbar\partial_\tau b+\hbar \clbar \partial_\tau \hatclk - \eabs \gbar_1 \gamma_1-\Omega\bbar b- \lambda \gbar_1 \gamma_1 (\bbar+b)-\chi(\bbar+b)- \sum_k\elk \clkbar \clk \biggr] \nonumber\\
&\qquad\qquad\qquad   -i\int_{-\infty}^{\infty}d\tau_1 \biggl\{  \frac{\ttilde}{\varepsilon_2}\biggl[(\utilde \clbar(\tau_1) + \vtilde^* \cl(\tau_1))(b+ \bbar)\bigl[(\lambdatc_{12})^*\gamma_1(\tau_1) + (\lambdatd_{12})^* \gbar_1(\tau_1) \bigr]\biggr] + \mathrm{H.c.}\biggr\},
\end{align}
where we have defined $\eabs \equiv \varepsilon_1$ and $\lambda \equiv 2 \lambdatc_{11}$. From Eq.~\eqref{eq:Zeff}, we can identify the effective Hamiltonian for  the tunnel coupling as
\begin{align}
\HT&= \frac{\ttilde}{\varepsilon_2}\biggl[ (\utilde \hatcl^\dagger + \vtilde^* \hatcl)(\hatb + \hatb^\dagger)\bigl[(\lambdatc_{12})^*\hatgamma + (\lambdatd_{12})^* \hatgamma^\dagger \bigr]\biggr] + \mathrm{H.c.}\nonumber\\
&= t(\hatb + \hatb^\dagger)\hatcl^\dagger(u\hatgamma +v\hatgamma^\dagger) + \mathrm{H.c.}
\end{align}
where we have defined $ \hatgamma \equiv \hatgamma_1$ as well as redefined
\begin{subequations}
\begin{align}
 t &\equiv \tilde{t} \frac{\tilde{\lambda}_{12}}{\varepsilon_2}, \label{eq:ttilde}\\
u &\equiv \utilde\frac{(\lambdatc_{12})^*}{\tilde{\lambda}_{12}} -\vtilde\frac{\lambdatd_{12}}{ \tilde{\lambda}_{12}},\label{eq:utilde}\\
  v &\equiv \utilde \frac{(\lambdatd_{12})^*}{\tilde{\lambda}_{12}} - \vtilde \frac{\lambdatc_{12}}{\tilde{\lambda}_{12}}\label{eq:vtilde},
\end{align}
\end{subequations}	
with $\tilde{\lambda}_{12} \equiv \sqrt{\left|\utilde(\lambdatc_{12})^* - \vtilde\lambdatd_{12}\right|^2 + \left|\utilde(\lambdatd_{12})^* - \vtilde\lambdatc_{12}\right|^2}$ which is chosen such that $|u|^2 + |v|^2 = 1$. So, the lead-ABS tunnel strength for the boson-assisted tunneling model is renormalized according to Eq.~\eqref{eq:ttilde}, and the particle-($u$) as well as the hole-component ($v$) of the ABS wave function seen by the electrons or holes tunneling from the lead are renormalized according to Eqs.~\eqref{eq:utilde} and~\eqref{eq:vtilde}, respectively. 
Note that the ABS Hamiltonian $\HA$ is the same as Eq.~\eqref{eq:HABSsuppl}.
By using the Lang-Firsov transformation as in Sec.~\ref{sec:LangFirsov}, we can eliminate the electron-boson interaction term from $\HA$. As a result, the tunneling Hamiltonian for the boson-assisted tunneling model transforms into
\begin{align}\label{eq:HTindirect}
\HTtilde &=   t (\hattildeb+\hattildeb^\dagger)\hatcl^\dagger \cat+\mathrm{H.c.},
\end{align}
where $\hattildeb$ and $\cat$ [Eq.~\eqref{eq:cantrans}] are the Lang-Firsov transformation of the operators $\hatb$ and $\ca = u\gamma + v\gamma^\dagger$.  

We now evaluate the matrix elements for the electron and hole tunneling which change the ABS occupancy number $n$ from $0 \rightarrow 1$ and the boson occupancy from $q \rightarrow q'$ using the Baker-Campbell-Haussdorf formula
\begin{equation}
\hatY^\dagger = e^{\frac{\lambda}{\Omega}(\hatb^\dagger - \hatb)}= e^{-\frac{\lambda^2}{2\Omega^2}}e^{\frac{\lambda}{\Omega} \hatb^\dagger}e^{-\frac{\lambda}{\Omega} \hatb},
\end{equation}
which gives
\begin{subequations}\label{eq:bpb}
\begin{align}
\langle 1,q'|(\hatb + \hatb^\dagger- 2\frac{\lambda}{\Omega}\hatgamma^\dagger \hatgamma)\hattildegamma^\dagger |0,q\rangle &= e^{-\frac{\lambda^2}{2\Omega^2}}\langle 1|\hatgamma^\dagger |0\rangle   \langle q'| e^{\frac{\lambda}{\Omega} \hatb^\dagger}(\hatb^\dagger + \hatb) e^{-\frac{\lambda}{\Omega} \hatb}|q\rangle-\frac{\lambda}{\Omega}\langle 1|\hatgamma^\dagger |0\rangle   \langle q'|\hatY^\dagger|q\rangle,\\
\langle 0,q'|\hattildegamma(\hatb + \hatb^\dagger- 2\frac{\lambda}{\Omega}\hatgamma^\dagger \hatgamma) |1,q\rangle &=\langle 1,q|(\hatb + \hatb^\dagger- 2\frac{\lambda}{\Omega}\hatgamma^\dagger \hatgamma)\hattildegamma^\dagger |0,q'\rangle^*.
\end{align}
\end{subequations}
The explicit expressions of Eq.~\eqref{eq:bpb} can be obtained from Eq.~\eqref{eq:Yqqp} and the following equations:
\begin{subequations}\label{eq:bpbelem}
\begin{align}
\langle q'|e^{\frac{\lambda}{\Omega} \hatb^\dagger}\hatb e^{-\frac{\lambda}{\Omega} \hatb}|q \rangle 
 &=  \sum_{m=0}^{\mathrm{min}(q',q-1)}\left(\frac{\lambda}{\Omega}\right)^{q'-m}\left(-\frac{\lambda}{\Omega}\right)^{q-m-1}\frac{\sqrt{q'!q!}}{m!(q'-m)!(q-m-1)!},\\
\langle q'|e^{\frac{\lambda}{\Omega} \hatb^\dagger}\hatb^\dagger e^{-\frac{\lambda}{\Omega} \hatb}|q \rangle 
 &=  \sum_{m=0}^{\mathrm{min}(q'-1,q)}\left(\frac{\lambda}{\Omega}\right)^{q'-m-1}\left(-\frac{\lambda}{\Omega}\right)^{q-m}\frac{\sqrt{q'!q!}}{m!(q'-m-1)!(q-m)!}.
\end{align}
\end{subequations}
In evaluating Eq.~\eqref{eq:bpbelem}, we have used Eq.~\eqref{eq:Yqqeq} and the following relations:
\begin{subequations}
\begin{align}
\hatb e^{-\frac{\lambda}{\Omega} \hatb}|q\rangle &= \sum_{m = 0}^\infty \frac{1}{m!} \left(-\frac{\lambda}{\Omega}\right)^m \hatb^{m+1}|q\rangle =\sum_{m =0}^{q-1} \frac{1}{m!}\left(-\frac{\lambda}{\Omega}\right)^m\sqrt{\frac{q!}{(q-m-1)!}} |q-m-1\rangle,\\
\langle q'| e^{\frac{\lambda}{\Omega} \hatb^\dagger} \hatb^\dagger&= \sum_{l = 0}^{\infty} \frac{1}{l!}\left(\frac{\lambda}{\Omega}\right)^l \langle q'|(\hatb^\dagger)^{l+1} = \sum_{l = 0}^{q'-1} \frac{1}{l!}\left(\frac{\lambda}{\Omega}\right)^l\sqrt{\frac{q'!}{(q'-l-1)!}}\langle q'-l-1|.
\end{align}
\end{subequations}

For the tunneling Hamiltonian in Eq.~\eqref{eq:HTindirect}, the rates of the boson-assisted electron and hole tunneling processes can be calculated from  Fermi's Golden Rule to be
\begin{subequations}\label{eq:R0esuppl}
\begin{align}
\Rne &= \frac{2 \pi t^2\nu_0}{\hbar}   \left|\left\langle \bar{n},q'\left| (\hattildeb+\hattildeb^\dagger) \cat^\dagger\right|n,q\right\rangle\right|^2 f(E_{\bar{n},q'} - E_{n,q} - eV),\\
\Rnh &= \frac{2 \pi t^2 \nu_0}{\hbar} \left|\left\langle \bar{n},q'\left| (\hattildeb+\hattildeb^\dagger) \cat \right|n,q\right\rangle\right|^2 f(E_{\bar{n},q'} - E_{n,q} + eV),
\end{align}
where $\langle \bar{n}|\ca^\dagger |n \rangle$ and $\langle \bar{n}|\ca |n \rangle$ are the bare tunneling matrix elements for electrons and holes, respectively, and $f(E) = [1+\mathrm{exp}({E/k_{\mathrm{B}}T})]^{-1}$ is the lead Fermi function.
\end{subequations}
Using Eqs.~\eqref{eq:bpb} and~\eqref{eq:bpbelem}, we can evaluate the rates as
\begin{subequations}
\begin{align}
R^{0\rightarrow 1;e}_{q\rightarrow q'}&=\frac{\Gamma|u|^2}{\hbar} \biggl|X_{qq'} - \frac{\lambda}{\Omega}Y_{qq'}\biggr|^2 f(E_{1,q'} - E_{0,q} - eV),\\
R^{0\rightarrow 1;h}_{q\rightarrow q'}&=\frac{\Gamma|v|^2}{\hbar} \biggl|X_{qq'} - \frac{\lambda}{\Omega}Y_{qq'}\biggr|^2 f(E_{1,q'} - E_{0,q} + eV),\\
R^{1\rightarrow 0;e}_{q\rightarrow q'}&=\frac{\Gamma|v|^2}{\hbar} \biggl|X_{q'q} - \frac{\lambda}{\Omega}Y_{q'q}\biggr|^2 f(E_{0,q'} - E_{1,q} - eV),\\
R^{1\rightarrow 0;h}_{q\rightarrow q'}&=\frac{\Gamma|u|^2}{\hbar} \biggl|X_{q'q} - \frac{\lambda}{\Omega}Y_{q'q}\biggr|^2 f(E_{0,q'} - E_{1,q} + eV),
\end{align}
\end{subequations} 
with the boson matrix elements given by
\begin{subequations}
\begin{align}
Y_{qq'} &\equiv\langle q'|\hatY^\dagger|q\rangle = \langle q'|e^{\lambda(\hatb^\dagger-\hatb)/\Omega}|q\rangle, \\
X_{qq'} &\equiv e^{-\frac{\lambda^2}{2\Omega^2}} \langle q'| e^{\frac{\lambda}{\Omega} \hatb^\dagger}(\hatb^\dagger + \hatb) e^{-\frac{\lambda}{\Omega} \hatb}|q\rangle, \label{eq:Xqq}
\end{align}
\end{subequations}
where the explicit expressions for $Y_{qq'}$ and $X_{qq'}$ can be obtained from Eqs.~\eqref{eq:Yqqp} and~\eqref{eq:bpbelem}, respectively. 

For the boson-assisted tunneling model, we can also show that the conductance is in general PH antisymmetric unless $|u| = |v|$. To this end, we replace $|Y_{qq'}|$ by $|X_{qq'} - \lambda Y_{qq'}/\Omega|$ in the derivation for the proof given in Sec.~\ref{sec:proof}. Furthermore, we note that in contrast to the tunneling into boson-coupled ABS model where the peak area of the conductance vs voltage curve is constant with temperature (see Sec.~\ref{sec:proofarea}), for the boson-assisted tunneling model [Eq.~\eqref{eq:HTindirect}] the conductance peak area increases with increasing temperature.

\end{document}